\documentclass[useAMS,preprint]{aastex}
\usepackage[dvips]{epsfig,color}
\usepackage{xr}
\usepackage{url}
\usepackage{subfigure}

\newcommand\sersic{S\'{e}rsic}

\def\kms{\ifmmode km\thinspace \mbox{s}^{-1}\else km\thinspace$\mbox{s}^{-1}$\fi}     
\def\deg{\ifmmode^\circ\else$^\circ$\fi}  
\def\arcs{\ifmmode {'' }\else $'' $\fi}  
\def\arcm{\ifmmode {' }\else $' $\fi}    

\def\msun{$M_\odot$}

\def\hi{\ion{H}{1}}
\def\h2{$\mbox{H}_2$}
\def\lapp{\ifmmode {_<\atop{^\sim}} \else {$_<\atop{^\sim}$}\fi} 
\def\gapp{\ifmmode {_>\atop{^\sim}} \else {$_>\atop{^\sim}$}\fi} 

\slugcomment{Submitted to ApJ}

\shorttitle{Effect of Halo Mass on Gas Removal}
\shortauthors{Yoon and Rosenberg}

\begin{document}

\title{The Effect of Halo Mass on the \hi\ Content of Galaxies in Groups and Clusters}

\author{Ilsang Yoon\altaffilmark{1} and Jessica L. Rosenberg}
\affil{Department of Physics and Astronomy, George Mason University, Fairfax, VA 22030}
\email{yoon@strw.leidenuniv.nl}
\altaffiltext{1}{Current Address: Leiden Observatory, Leiden University, 2300 RA Leiden, The Netherlands\\}

\begin{abstract}
We combine data from the Sloan Digital Sky Survey (SDSS) and the Arecibo Legacy Fast
ALFA Survey (ALFALFA) to study the cold atomic gas content of galaxies in groups and
clusters in local  universe. A careful cross-matching of galaxies in the SDSS,
ALFALFA and SDSS group catalogs provides a sample of group galaxies with stellar
masses $10^{8.4} M_{\odot} \le M_{*} \le 10^{10.6} M_{\odot}$ 
and group halo masses $10^{12.5} h^{-1} M_{\odot} \le M_h \le 10^{15.0} h^{-1} M_{\odot}$.
Controlling our sample in stellar mass and redshift, we find no significant
radial variation in the galaxy \hi\ gas-to-stellar mass ratio for the halo mass range 
in our sample. However, the fraction of galaxies detected in ALFALFA declines steadily 
towards the centers of groups with the effect being most prominent in the most massive
halos. In the outskirts of massive halos a hint of a depressed detection fraction for 
low mass galaxies suggests pre-processing that decreases the \hi\ in these galaxies before 
they fall into massive clusters. We interpret the decline in the ALFALFA detection of galaxies 
in the context of a threshold halo mass for ram pressure stripping for a given galaxy
stellar mass. The lack of an observable decrease in the galaxy \hi\ gas-to-stellar mass ratio 
with the position of galaxies within groups and clusters highlights the difficulty of detecting 
the impact of environment on the galaxy \hi\ content in a shallow \hi\ survey. 
\end{abstract}


\section{INTRODUCTION}

The amount of cold gas in group and cluster galaxies provides information on the impact of 
these environments on galaxy evolution. Previous studies have shown that the distribution of cold gas in galaxies,
traced by neutral hydrogen (\hi), is affected primarily by tidal interactions \citep{toomre_1972} and 
ram pressure stripping \citep{gunn_and_gott_1972} (see \citealp{boselli_and_gavazzi_2006} for review of 
physical processes in galaxy environment).

Tidal interactions are expected to be important in galaxy groups but less so in clusters due to the 
large relative velocities of galaxies and short interaction time scales \citep{mihos2004}. Alternatively, 
ram pressure stripping is thought to be responsible for \hi-deficiency in cluster 
galaxies \citep[][]{boselli_and_gavazzi_2006,roediger_2009,gavazzi_etal_2013a}. Furthermore, ram pressure 
stripping can be felt by any galaxy containing gas. Ram pressure can strip the cold gas in ellipticals 
\citep[e.g.,][]{lucero_etal_2005,mccarthy_etal_2008} and dwarfs \citep[e.g.,][]{mcconnachie_etal_2007,freeland_and_wilcots_2011}, 
and the hot halo gas in galaxies \citep[e.g.,][]{jeltema_etal_2008} causing `starvation' \citep{larson_etal_1980}. 
For this work we select our sample to minimize the impact of tidal interactions so that we can focus on the effects of 
ram pressure stripping.

The impact of ram pressure stripping on galaxies has been well studied in the Virgo and Coma clusters. 
Statistical analyses of the \hi\ gas content of their member galaxies provides evidence for increasing gas depletion 
toward the cluster center for optically-selected galaxies \citep[e.g.,][]{haynes_etal_1984,haynes_etal_1986,solanes_etal_2001} 
and for \hi-selected late type galaxies from ALFALFA \citep[][]{gavazzi_etal_2013a}. The disturbed \hi\ 
morphologies of individual galaxies in Virgo and Coma cluster have provided additional evidence for on-going ram pressure 
stripping \citep[e.g.,][]{bravo_etal_2000,bravo_etal_2001,chung_etal_2009,chung_etal_2007,crowl_etal_2005,kenney_etal_2004,vollmer_etal_2008}. 
Nevertheless, \citet{taylor_etal_2012} and \citet{cortese_etal_2008} 
find only marginal evidence for increasing \hi-deficiency toward the inner regions of the Virgo and the Coma/A1367 
clusters respectively using \hi-selected samples from the Arecibo Galaxy Environment Survey \citep{auld_etal_2006}. 
This lack of \hi\ depletion in these clusters is likely due to the limited dynamic range of this flux-limited \hi\ survey. 

While statistical and individual observations suggest that interaction between galaxies and the 
intracluster medium is important for the removal of gas from galaxies in the most massive groups, the situation 
is less clear in smaller groups \citep{rasmussen_etal_2012}. In a small number of low mass groups there is evidence for 
individual galaxies that have been ram-pressure stripped of their gas 
\citep[e.g.,][]{bureau_and_carignan_2002,freeland_and_wilcots_2011,mcconnachie_etal_2007,rasmussen_etal_2006,sengupta_etal_2007}.
Some of these groups also have X-ray detections of a hot intergalactic medium (IGM) consistent with the ram pressure stripping explanation. 

A few recent studies have examined the impact of environment for wide range of galaxy group halo 
masses using statistical sample. \cite{catinella_etal_2013} used $\approx 800$ galaxies from the GALEX Arecibo SDSS survey 
(GASS; \citealp{catinella_etal_2010,catinella_etal_2012}) to show that the \hi\ gas fraction 
(defined as $M_{\mbox{\tiny HI}}/M_{*}$ in their work) is at least 0.4 dex smaller for galaxies with stellar masses 
$M_{*}\ge10^{10.0} M_{\odot}$ that reside in groups with halo mass $M_h>10^{13-14} M_{\odot}$. \cite{fabello_etal_2012} stacked ALFALFA 
spectra of galaxies meeting the GASS selection criteria in bins of stellar mass and local density and found a strong decline in the 
galaxy \hi\ content in dark matter halos with mass $M_h>10^{13} M_{\odot}$. \cite{hess_and_wilcots_2013} used the SDSS group 
catalog \citep{berlind_etal_2006} and the ALFALFA 40\% catalog \citep{haynes_etal_2011} to show that as group membership increases, 
galaxy group centers become increasingly deficient of \hi-rich galaxies. 
This work builds on these studies by using a control sample (i.e., isolated field galaxies with similar stellar mass and redshift) 
to compare the impact of the groups and clusters to the \hi\ gas content of galaxies.

Hydrodynamic simulations show that ram pressure stripping can remove the outer  \hi\ gas from a galaxy falling into cluster
\citep[e.g.,][]{abadi_etal_1999,marcolini_etal_2003,roediger_and_bruggen_2007,tonnessen_and_bryan_2009,tonnessen_and_bryan_2010}
within 10-100 Myr  \citep{abadi_etal_1999,marcolini_etal_2003,roediger_2009,tecce_etal_2010} while complete unbinding of 
the disk gas may take a few 100 Myr \citep{roediger_2009}, less than the dynamical time scale in groups 
and clusters ($\approx 1$ Gyr, \citealp{boselli_and_gavazzi_2006}). 
Theoretical models of ram pressure stripping based on the original formulation in \citet{gunn_and_gott_1972} find that 
the ratio of the mass of the infalling galaxy to that of the host group or cluster is an important parameter governing the 
efficiency of ram pressure stripping \citep{hester_2006} and the fraction of galaxies without cold gas increases toward the 
cluster center due to increasing ram pressure \citep{tecce_etal_2010}. Therefore one can expect that
gas depletion will increase toward the center of groups and clusters and that the depletion will be more significant 
in more massive clusters.

We use a statistical sample of galaxies with \hi\ measurements from the ALFALFA survey to investigate the 
distribution of galaxy \hi\ gas-to-stellar mass ratio as a function of the projected distance from the center of galaxy group halos 
for wide range of halo masses. We have combined the ALFALFA \hi\ source catalog, the SDSS photometric and 
spectroscopic catalog, and the SDSS group catalog. In addition, we have created a control sample to remove the biases that arise
from the correlation between stellar mass and \hi\ gas-to-stellar mass ratio \citep{catinella_etal_2012,huang_etal_2012}.

In Section \ref{data}, we introduce the data sets used in this study and discuss our sample selection. 
In Section \ref{sample} we describe the  procedure used to match SDSS galaxies, group catalog, and \hi\ 
sources from ALFALFA. In Section \ref{result}, we examine the distribution of galaxy \hi\ gas-to-stellar mass ratio and 
the fraction of the \hi\ detected galaxies as a function of the projected distance from the group center and derive a simple ram pressure 
stripping criterion depending on group halo mass which provides a plausible explanation to our findings. 
In Section \ref{discussion} we discuss the implications of our results and the relevant 
issues and summarize the results in Section \ref{conclusion}. If not explicitly stated otherwise, we use a spatially flat LCDM 
cosmology with $\Omega_m=0.3$, $\Omega_{\Lambda}=0.7$ and $H_0=100 h$ $\mbox{km s}^{-1}\mbox{Mpc}^{-1}$, where $h=0.73$.  

\section{DATA}\label{data}

The galaxies in this sample are selected from the NASA Sloan Atlas catalog \footnote{\url{http://www.nsatlas.org/}} \citep{blanton_etal_2011} 
and have \hi\ 21 cm measurements in the ALFALFA 40\% catalog \citep[][]{haynes_etal_2011} and a group identification in the SDSS group 
catalog \citep{yang_etal_2007}. We restrict the sample to galaxies in halos with masses 
$10^{12.5} h^{-1} M_{\odot} \le M_h \le 10^{15.0} h^{-1} M_{\odot}$, more than 4 members, and redshifts of $0.01 < z < 0.055$. 
A complementary sample of isolated galaxies is used to create a control sample for this study.  The cross-matching of the 
three catalogs is discussed in detail in Section \ref{sample} but it is useful to note that the catalogs have very different selection 
functions and completeness. Requiring that the galaxies be in all three catalogs 
produces a sample that is mainly limited by the ALFALFA selection function.

\subsection{The Optical Galaxy Sample} \label{data:sdss}

Our optical galaxy sample is selected from the NASA Sloan Atlas catalog \citep{blanton_etal_2011} which includes photometric and spectroscopic 
measurements for $\sim$150,000 galaxies within $\approx 220$ Mpc ($z \lapp 0.055$) selected from SDSS DR8 \citep{aihara_etal_2011}. The size of 
the NASA Sloan Atlas catalog is much smaller than the full SDSS DR8 because of the redshift limit. 
However, the NASA Sloan Atlas 
catalog provides more reliable and accurate photometry than the SDSS DR8 catalog. In particular, the improved background subtraction removes the 
size-dependent flux bias for galaxies with half-light radii of $r_{50} \le 100$\arcsec \citep{blanton_etal_2011}. The NASA Sloan 
Atlas catalog also provides an improved photometric center for galaxies in which the SDSS spectrum is not coincident with the photometric 
center and has eliminated objects where the SDSS spectroscopic object is a sub-clump of a larger galaxy. These improvements allow for 
a more reliable estimate of stellar mass and improved matching with the ALFALFA catalog galaxies. 

The optical photometry from the NASA Sloan Atlas catalog is used to derive the stellar masses of the galaxies. 
Galaxy stellar mass is computed from galaxy color and luminosity following \citet{bell_etal_2003}:

\begin{equation}
\label{eq:mstar}
\mbox{log} \left(\frac{M_{*}}{h^{-2} M_{\odot}}\right) = -0.306+1.097(g-r)-0.1-0.4(M_r-5\mbox{log}h-4.64)
\end{equation} 
where $M_r$ is the SDSS $r$-band absolute magnitude and $(g-r)$ is the $g-r$ color based on the \sersic\ model magnitude. 
The color-based stellar mass measurement is used for consistency with the stellar mass definition in the SDSS group 
catalog \citep{yang_etal_2007}. However, the results presented here do not change if stellar masses derived from SED fitting are 
used instead.  

\subsection{The \hi\ Galaxy Sample} \label{data:alfalfa}

The ALFALFA survey is an extragalactic \hi\ 21 cm line survey that provides information about the neutral atomic gas content of 
nearby galaxies ($z<0.06$, \citealp{giovanelli_etal_2005}). The survey covers $\sim$7000 square degrees of sky with 
10 $\mbox{km s}^{-1}$ velocity resolution (after Hanning smoothing), an rms sensitivity of 1.8 mJy beam$^{-1}$ \citep{giovanelli_etal_2005}, 
and a positional accuracy of $\sim$24\arcsec\ for sources with $\mbox{S/N}>6.5$ \citep{giovanelli_etal_2007}. The currently 
available source catalog, $\alpha.40$, contains $\approx 15000$ sources within 40\% of the total survey area \citep[][]{haynes_etal_2011} 
and includes measurements of redshift, \hi\ flux, and \hi\ line width that can 
be used to determine the \hi\ mass of the source using the relation \citep{haynes_etal_2011}: 

\begin{equation}
M_{\mbox{\tiny HI}} = 2.36 \times 10^5 D^2 \int S dv \quad M_{\odot}
\end{equation}
where $D$ is the distance to the galaxy in Mpc, 
$S$ is the \hi\ 21 cm flux density, and $\int Sdv$ is the \hi\ flux of the source integrated over the velocity. 

Both ``code 1" sources, which are highly reliable detections with $\mbox{S/N}>6.5$, and ``code 2" sources, which are lower 
signal-to-noise detections ($4.5<\mbox{S/N}<6.5$) that are likely to be real because there is a known optical counterpart at the 
redshift of the \hi\ source \citep{haynes_etal_2011}, are included in our sample. The majority (76\%) of the sample consists 
of ``code 1" sources and the fraction of ``code 2" sources does not vary significantly with the projected distance from 
group center. 

\begin{figure}
\includegraphics[]{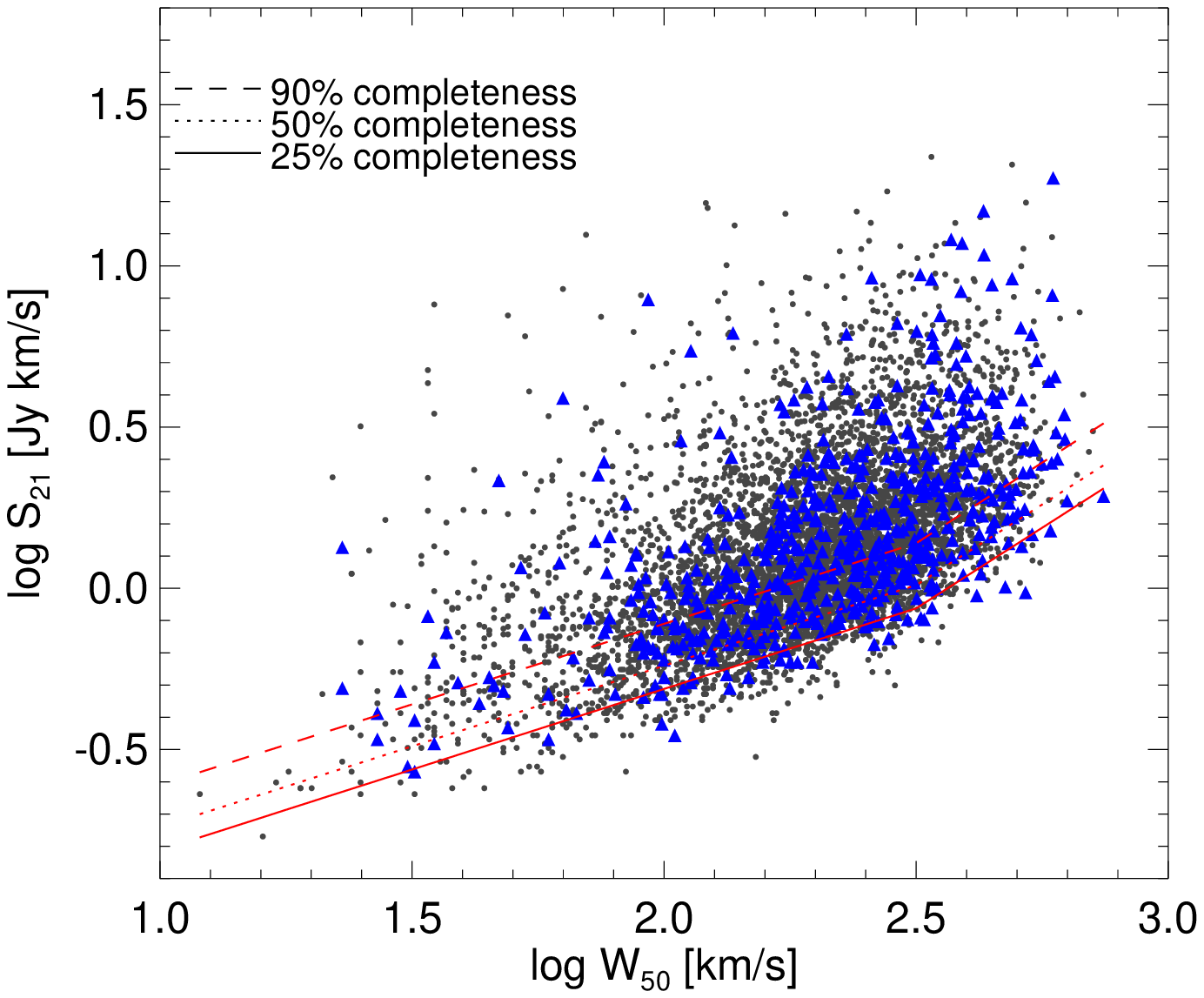}
\caption{Distribution of the galaxy sample from cross-matching between the NASA Sloan Atlas, the SDSS group and the $\alpha.40$ catalog 
in the \hi\ flux and line width plane. Triangles (blue in the online version) are the group samples and grey dots are the control samples.
Lines (red in the online version) are completeness limits determined by \cite{haynes_etal_2011}. For both the group sample and the control sample
approximately 41\% and 6\% of galaxies are under the 90\% and 25\% completeness limits respectively.}\label{fig:completeness}
\end{figure}

Figure \ref{fig:completeness} shows the relationship between the velocity integrated \hi\ flux and the \hi\ 21 cm line width measured at 50\% of 
the peak flux, for the sample in this study. The 25\%, 50\%, and 90\% completeness limits for the $\alpha.40$ catalog \citep{haynes_etal_2011} 
are indicated by the red lines in the figure. The blue triangles are the group galaxies 
and the grey points are the isolated galaxies used as controls (see \S \ref{sample:xmatch} for details on how this sample 
was selected from the cross-match of the three catalogs). The fraction of the sample below each completeness limit is nearly same 
for the isolated and group samples: 41\% and 6\% of them are below the 90\% and 25\% completeness limit. 

\subsection{The Galaxy Group Catalog} \label{data:group}

The goal of this study is to understand the impact of the group environment on the \hi\ gas content in galaxies. 
To determine the group membership of galaxies in our sample, we use the SDSS DR7 group 
catalog\footnote{\url{http://gax.shao.ac.cn/data/Group.html}} updated from the DR4 group catalog \citep{yang_etal_2007}. 
The catalog uses galaxies in the SDSS DR7 spectroscopic sample with $0.01 \le z \le 0.2$ 
and redshift completeness $C>0.7$. The group finding algorithm has been extensively tested using mock galaxy redshift survey 
catalogs and has proven to be successful in associating galaxies that reside in a common halo \citep{yang_etal_2005}. 
In particular, this halo-based group finder works well for poor groups and identifies groups with only one member 
(i.e., isolated galaxies). The group halo masses are determined down to $M_h=10^{11.8} h^{-1} M_{\odot}$ using two methods: ranking 
by luminosity and stellar mass of galaxies in groups. Although we used the luminosity ranked group halo mass, the results do not 
change if the stellar mass ranked halo mass is used instead. The group finder has been shown to correctly select more than 90\% of the true 
halos with $M_h \ge 10^{12} h^{-1} M_{\odot}$ \citep{yang_etal_2007}, which allows us to reliably study our galaxy samples within 
groups and clusters with halo mass $10^{12.5} h^{-1} M_{\odot} \le M_h \le 10^{15} h^{-1} M_{\odot}$. 

In order to investigate galaxy \hi\ properties as a function of the projected distance from the group center, we normalize each 
galaxy's radial position by the virial radius of the group. For virial radius of group with halo mass 
$M_h$, we adopt the radius $R_{180}$ that encloses an overdensity of 180 relative to the critical density \citep{yang_etal_2007}: 

\begin{equation}
\label{eq:rvir}
R_{180}=1.26 h^{-1} \mbox{Mpc} \left(\frac{M_h}{10^{14}h^{-1} M_{\odot}}\right)^{1/3} (1+z_{group})^{-1}
\end{equation} 
which is based on the WMAP3 cosmological model 
parameters \citep{spergel_etal_2007}, $\Omega_m=0.238$, $\Omega_{\Lambda}=0.762$ and $H_0=100 h$ $\mbox{km s}^{-1}\mbox{Mpc}^{-1}$, 
where $h=0.73$. While these parameters differ slightly from those used in this study, using this formulation does not make a significant 
difference given the low redshifts of our sample ($0.01<z<0.055$).

\begin{figure}
\centering
\subfigure[]{\epsfig{figure=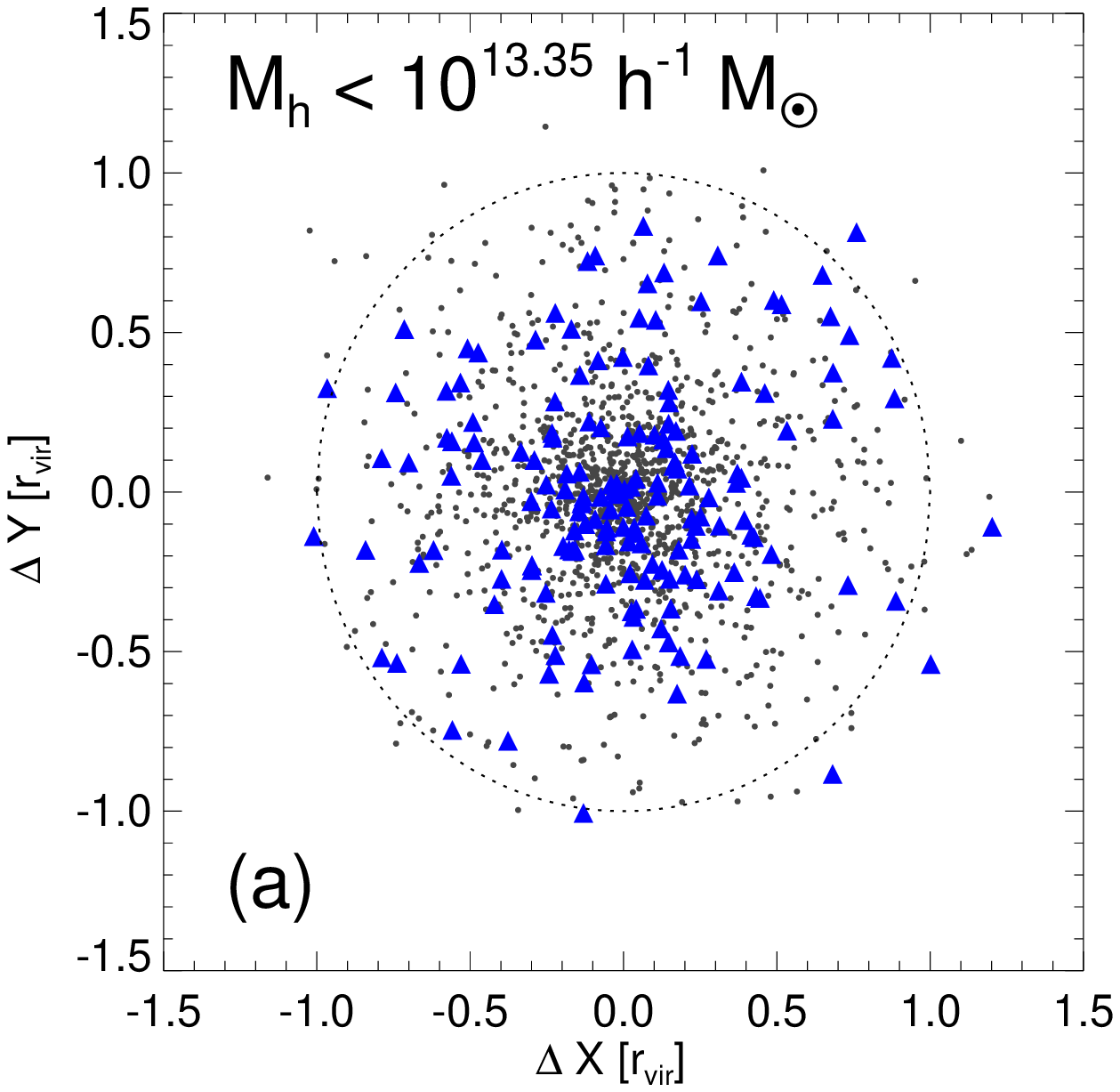, scale=0.5}}
\subfigure[]{\epsfig{figure=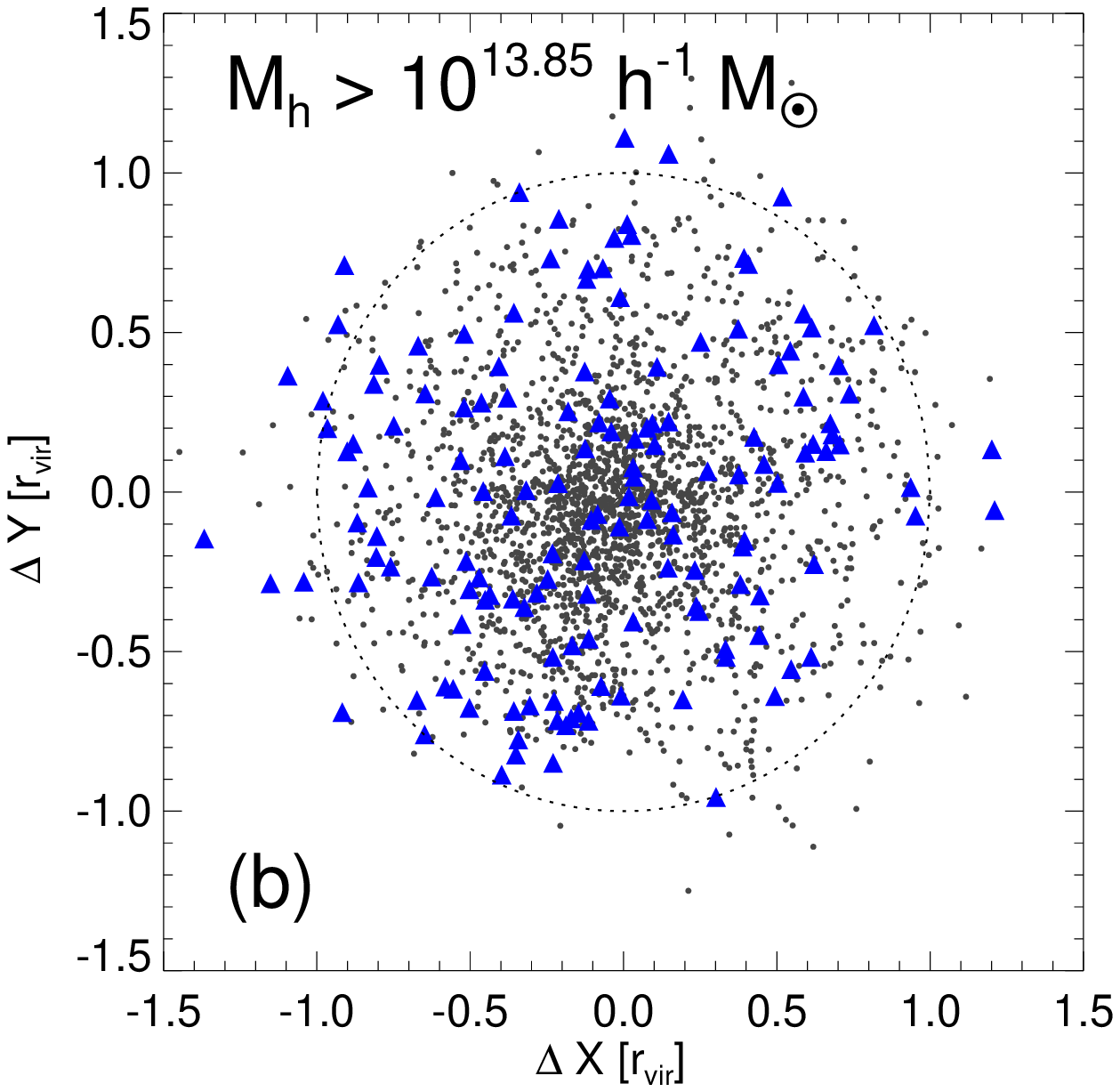, scale=0.5}}
\caption{Stacked position of group member galaxies relative to the center, normalized by each group halo's $R_{180}$.
The grey dots are the SDSS galaxies within the $\alpha.40$ catalog footprint and triangles (blue in the online version) are the ALFALFA detected 
galaxies. The dotted line represents $R_{180}$. 
Left: galaxies in groups with $M_h < 10^{13.35} h^{-1} M_{\odot}$. 
Right: galaxies in groups with $M_h > 10^{13.85} h^{-1} M_{\odot}$}\label{fig:samplepos}
\end{figure}

Figure \ref{fig:samplepos} shows the normalized positions of the galaxies in the most massive and least massive group halos 
($M_{h} < 10^{13.35} h^{-1} M_{\odot}$ and $M_{h} > 10^{13.85} h^{-1} M_{\odot}$) in this study. Complete details regarding the galaxy samples in
groups are discussed in Section \ref{sample:group}. Compared with the distribution of the optically-selected galaxies (grey dots), 
the ALFALFA detected galaxies (blue triangles) are rarer near the centers of the groups with the effect significantly more pronounced 
for the more massive groups. This lack of \hi\ detections near group centers matches the findings of \citet{hess_and_wilcots_2013}. 
In Section \ref{result:detection} we investigate the \hi\ detection fraction more carefully by using a control sample to avoid the
effect of \hi\ selection bias by the masses and distances of galaxies at different positions within group halos.

\section{CROSS-MATCHING AND SAMPLE DEFINITION}\label{sample}

\subsection{Cross-matching SDSS, ALFALFA, and the Group Catalog} \label{sample:xmatch}

For each SDSS galaxy in the NASA Sloan Atlas catalog, the associated group halo was identified by locating 
\footnote{All matching was done using the catalog manipulation software \textit{Topcat} \citep{taylor_2005}} 
the SDSS galaxy in the group catalog that is within 5\arcsec\ in position. This produces 103,287 SDSS galaxy matches. 
Since galaxies in the SDSS group catalog and in the NASA Sloan Atlas catalog are the same sources, 5\arcsec\ positional matching 
is very robust and the redshift difference is typically $\Delta z \approx 0.0001$ (i.e., 30 \kms\ in velocity). Only 0.1\% of 
matches have redshift differences of $0.0003 < \Delta z < 0.0009$ (none have $\Delta z > 0.0009$). 
The larger velocity differences occur for galaxies where the NASA Sloan Atlas Catalog uses a non-SDSS redshift measurement that 
is deemed more reliable. Generally this happens when SDSS does not have a spectrum near the photometric center of the galaxy.
All 103,287 NASA Sloan Atlas galaxies that are associated with groups were then cross-matched with the $\alpha.40$ catalog. 
Sources within 40\arcsec\ in position and 0.001 in redshift (300 \kms\ in velocity) were considered potential matches. 

\begin{figure}
\centering
\includegraphics[]{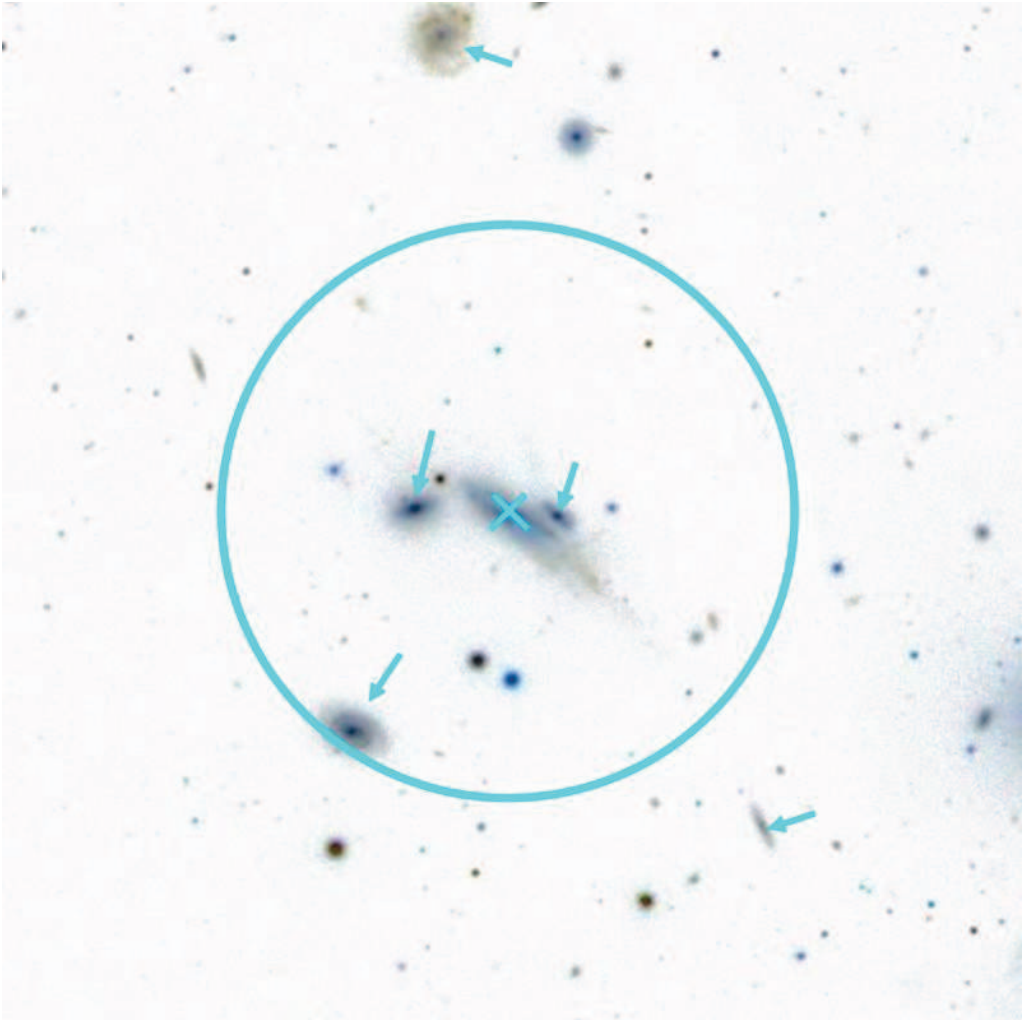}
\caption{Image of AGC8842 with Hickson Compact Group 69 galaxies within 1.75 \arcmin\ marked with arrows. The ALFALFA beam, shown as the blue 
circle here, includes multiple galaxies which boosts the gas-to-stellar mass ratio of the central galaxy identified as the best candidate for 
the ALFALFA source. We avoid such cases by removing groups that have ALFALFA sources that can be associated with multiple galaxies 
within a 1.75 \arcmin\ radius.}\label{fig:AGC8842}
\end{figure}

In total there were 6444 potential matches between SDSS group galaxies and ALFALFA sources.
However, since the Arecibo beam is large ($\sim$3.5\arcmin diameter), there can be more than one galaxy in the beam that will 
influence the measurement of the \hi\ mass. Figure \ref{fig:AGC8842} shows an extreme example in which there are 4 galaxies in 
Hickson Compact Group 69 within the ALFALFA beam. In this case, the \hi\ measurement is likely not that of a single galaxy. 
Cases of galaxy confusion will be more common in groups than in the field and thus the \hi\ mass of the best matched galaxy among the 
multiple galaxies within the Arecibo beam will be biased (e.g., the best match indicated by the cross in Figure \ref{fig:AGC8842}). 
To address this problem we exclude from the matched sample 520 galaxies that have additional SDSS sources 
within a 1.75\arcmin\ radius. Removing these galaxies with close neighbors decreases our sample size to 5924 (a decrease of 8\%) 
and removes systems that are likely to be experiencing strong and complex gravitational interactions 
as is the case in compact groups \citep[e.g.,][]{cluver_etal_2013,rasmussen_etal_2012} rather than purely galaxy-IGM 
interactions. This proximity limit excludes galaxies that are within 20 kpc at $z=0.01$ and 105 kpc at $z=0.055$.

The sample selected from matching the three different catalogs, excluding galaxies with close neighbors includes 5924 galaxies. 
For these galaxies, the catalogs provide SDSS galaxy parameters, SDSS group parameters, and the ALFALFA \hi\ 21 cm parameters. 
The lower limit on the redshift ($z=0.01$) of 
the cross-matched sample is set by the SDSS group catalog and the upper limit ($z=0.055$) is set by the NASA Sloan Atlas catalog.
This is the matched sample catalog from which we construct the final group sample with controls as discussed in the following section.

\subsection{Defining the Group and Control Samples} \label{sample:group}

Our cross-matched galaxy sample is limited primarily by the ALFALFA detection limits thus biasing the sample towards gas-rich systems. 
In particular, the detectability of an \hi\ source depends on the galaxy redshift and the \hi\ mass which is, at least loosely, 
correlated with the stellar mass \citep[e.g.,][]{catinella_etal_2010,catinella_etal_2012,huang_etal_2012}. To mitigate these biases, 
we constructed a control sample of isolated field galaxies. The control galaxies were selected to have similar stellar masses 
($\Delta \log M_{*} < 0.1$) and redshifts ($\Delta z < 0.001$) as their group counterparts. 

The catalog of 5924 galaxies provides the sample from which we drew both our isolated and group galaxies. Group galaxies were 
selected to have halo masses larger than $10^{12.5} h^{-1} M_{\odot}$, significantly larger than the 90\% completeness 
limit ($10^{12.0} h^{-1} M_{\odot}$) in the group catalog \citep{yang_etal_2007}, and group membership of more than 4 galaxies. 
Isolated galaxies were identified as galaxies residing in groups with one member. The number of galaxies with one group member (i.e., 
isolated field galaxies) and with group membership of more than 4 galaxies are 4372 and 618 respectively. The remaining 934 
galaxies which are not included in this study reside in groups with 2-4 members.  

\hi\ gas-to-stellar mass ratio is known to be correlated with galaxy stellar mass down to $M_{*}=10^{10} M_{\odot}$ with a large scatter 
(0.43 dex, \citealp{catinella_etal_2012}). This trend may continue to the lower masses but is harder to quantify because of sample 
incompleteness (\citealp{cortese_etal_2011,huang_etal_2012}; Rosenberg et al. in preparation). Because of the large scatter in 
this relationship, a small control sample will introduce a sampling bias and varying the numbers of controls
will produce differences in the sample variance. Therefore, we use the same number of control galaxies, 8, for every group galaxy.  

\begin{figure}
\centering
\epsfig{figure=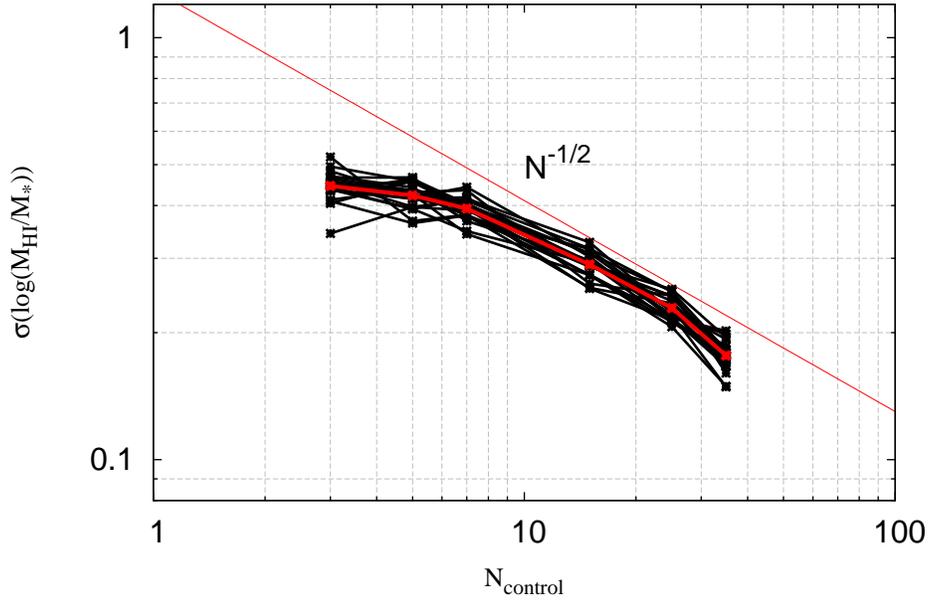, scale=0.5,angle=-90}
\caption{Standard deviation of the average $\mbox{log} (M_{\mbox{\tiny HI}}/M_{*})$ for the control samples for each pair of the 50 stellar 
mass and redshift values, as 
function of the number of control samples. The different black lines are 20 separate calculations of the above values for standard deviation 
as a function of control sample size. The grey (red in the online version) line is the `average' of these 20 lines. The thin solid (red in 
the online version) line has a slope of $1/\sqrt{n}$, the statistically expected change in the standard deviation as the number of controls 
change. The line has been placed with an arbitrary offset to show that the data matches the expected slope well for samples with 7 or more 
controls. However, for large numbers of controls the number of groups that have a full complement of controls drops rapidly. 
We find that the optimal control sample size is 8.}\label{fig:sigmatest}
\end{figure}

Forcing each group galaxy to have the same number of controls means that for a larger number of controls, the number of group galaxies 
with enough controls that meet the selection criteria ($\Delta \log M_{*} < 0.1$ and $\Delta z < 0.001$) decreases.
Alternatively, decreasing the number of control galaxies increases the number of group galaxies with a full complement of controls, 
but it also increases the variance in the control sample. To optimize the control sample size, we investigated how the standard 
deviation of the \hi\ gas-to-stellar mass ratio changes as the size of the control sample changes.
For 50 randomly selected pairs of values for stellar mass and redshift that lie within the survey range a control sample was assembled. 
The average $\log (M_{\mbox{\tiny HI}}/M_{*})$ for each group of controls was calculated, 
resulting in 50 values of this parameter. The mean and standard deviation of these 50 values was then computed. This procedure was 
repeated for control samples with $N = 3, 5, 7, 15, 25,$ and $30$ and the values for the standard deviation were plotted as a function of 
the number of controls in Figure \ref{fig:sigmatest}. The same computation was carried out 20 times and the results are shown by the 20 black lines. 
The red line is the average of the 20 trials. This figure shows that small control samples deviate significantly from the expected statistical 
behavior of variance with sample size (shown by the thin solid line), $n$, which is proportional to $1/\sqrt{n}$, because the controls 
are not representative of the \hi\ gas-to-stellar mass ratio distribution. Alternatively, as the number of controls increases, the number of available group 
galaxies declines.  

Figure \ref{fig:sigmatest} indicates that the optimal control sample size where the variance goes as $1/\sqrt{n}$ while maximizing the number of group 
galaxies is between 7 and 10. We use 8 control galaxies for each group 
galaxy as a balance between maximizing the number of controls and maximizing the number of group galaxies. Group galaxies with fewer than 
8 controls meeting the selection criteria ($\Delta M_{*} < 0.1$ dex and $\Delta z < 0.001$) are not included in the final sample. 
The best controls were selected by ranking the isolated galaxies based on the normalized distance from the group 
galaxy in the stellar mass-redshift plane (i.e., d$_{norm} = \sqrt{(\Delta \log M_{*}/0.1)^2 + (\Delta z/0.001)^2}$). 
The 8 closest isolated galaxies were selected as the control sample for each group galaxy. This technique is similar to 
that used by \cite{ellison_etal_2008}.

\begin{figure}
\centering
\epsfig{figure=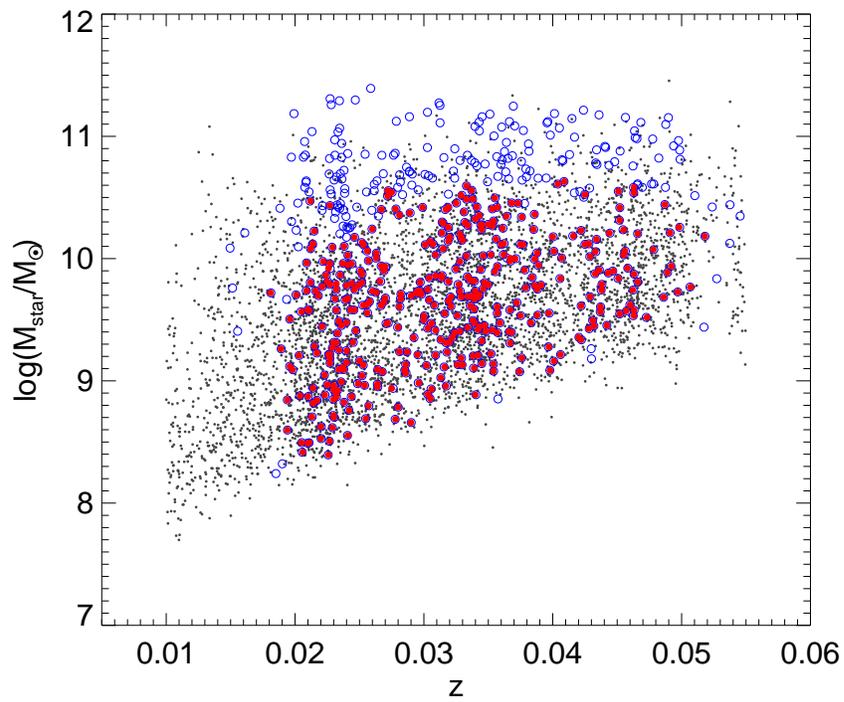, scale=0.8}
\caption{Stellar mass and redshift for galaxies in groups represented by circles (blue
in the online version) and for isolated galaxies represented by small black points. The filled circles 
(red in the online version) are the galaxies in the final sample that have a full complement of 8 controls.}
\label{fig:sampledist1}
\end{figure}

Figure \ref{fig:sampledist1} shows the distribution of stellar masses and redshifts for galaxies in the group and isolated samples. 
Small dots are the isolated galaxies used as controls and blue circles are the group galaxies. The filled circles show the 
final galaxy sample (390 group galaxies that each have 8 controls). Note that there are not many group galaxies with a full complement 
of controls above $M_{*} = 10^{10.63} M_{\odot}$. For this reason, our final sample is limited by $M_{*} < 10^{10.63} M_{\odot}$. 
The stellar mass and redshift distributions of the galaxy group sample and the sample of their associated controls are drawn from 
the same population with more than 99\% and 95\% confidence respectively, based on the Kolmogorov-Smirnov (K-S) test \citep{press_etal_1986}. 

\begin{figure}
\centering
\epsfig{figure=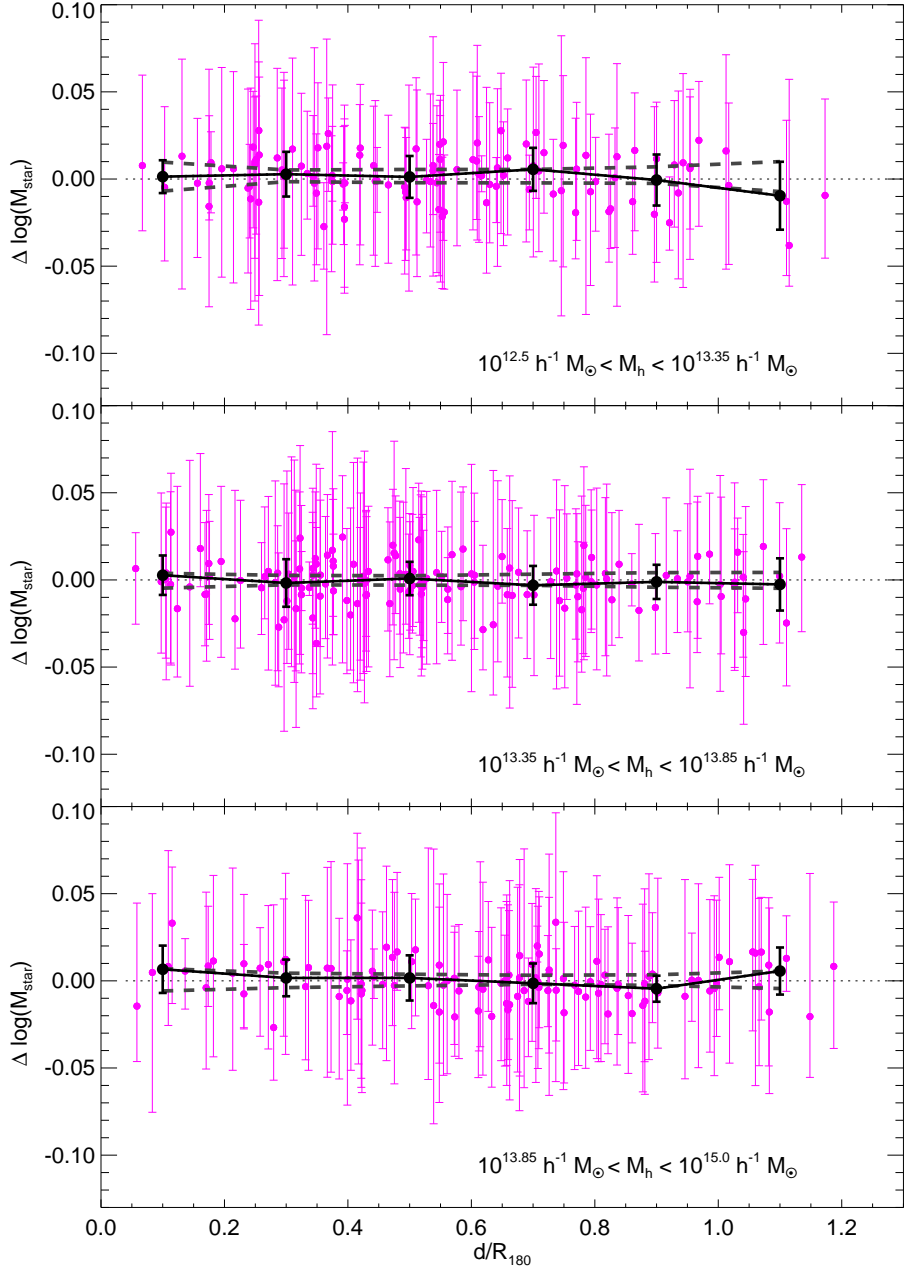, scale=0.70}
\vspace{0.3cm}
\caption{The difference between $\log M_{*}$ for the galaxies in the group sample and the log of the 
average $M_{*}$ for their 8 associated control galaxies as a function of the projected group centric distance for different halo mass bins. 
Projected distance is normalized by group virial radius ($R_{180}$). Data points (magenta in the online version) 
are individual galaxies and the associated error bar is the range of the value determined by the 8 control samples. The solid line represents 
a mean of the $\Delta \log M_{*}$ in each radial bin. The dashed lines enclose the $5^{th}$ and $95^{th}$
quantile of the radial distribution generated by random shuffling of galaxy radial position.
Note that no significant radial variation of the stellar mass relative to the controls ($\Delta \log M_{*} < 0.01$) 
is seen for galaxies in all halo mass bins.}
\label{fig:mstar_dist}
\end{figure}

\begin{figure}
\centering
\epsfig{figure=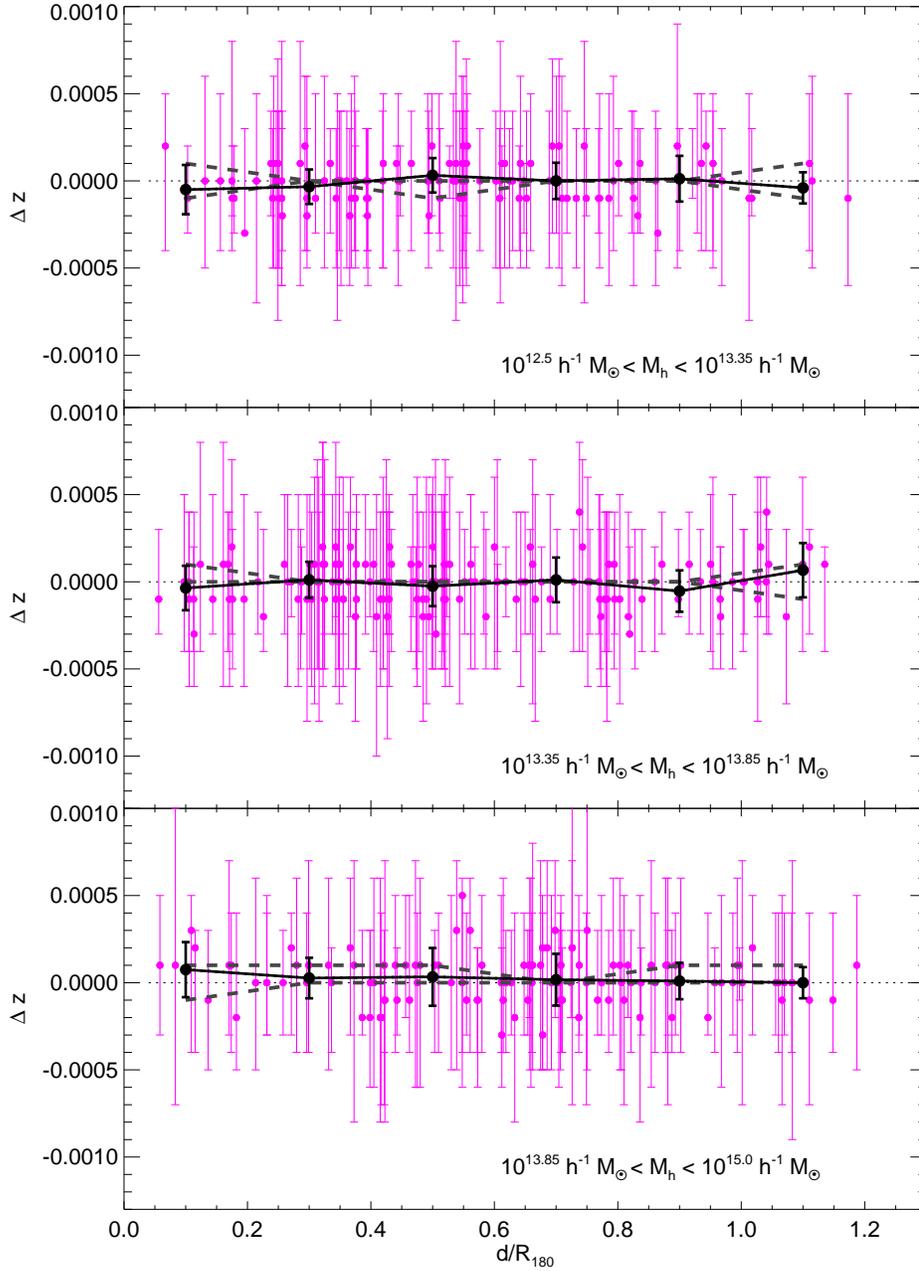, scale=0.70}
\vspace{0.3cm}
\caption{Same as Figure~\ref{fig:mstar_dist} for the difference of redshift between the galaxies in the group
sample and the mean of the corresponding 8 control galaxies, as a function of the projected group centric distance for different halo mass bins. 
As in Figure~\ref{fig:mstar_dist}, no significant radial variation of the redshift relative to the controls ($\Delta z < 0.0002$) 
is seen for galaxies in all halo mass bins.}
\label{fig:redshift_dist}
\end{figure}

Figure \ref{fig:mstar_dist} and \ref{fig:redshift_dist} show galaxy stellar mass and redshift for the group and control samples as a 
function of the projected distance from the 
group center normalized by the group virial radius ($R_{180}$). The three panels show different halo mass bins.
The $\Delta \log M_{*}$ values are the difference between the log of the stellar mass of the group galaxy 
and the log of the average stellar mass of its 8 controls. The $\Delta z$ values are the difference between the redshift of the group 
galaxy and the average redshift of its 8 controls. Error bars associated with each group galaxy correspond to the range of 
stellar mass and redshift spanned by the galaxy's 8 control samples. The black solid lines connect the average of $\Delta \mbox{log} M_{*}$ 
and $\Delta z$ in 6 radial bins up to 1.2 $R_{180}$ with errors defined by standard deviations of $\Delta \mbox{log} M_{*}$
and $\Delta z$ in each radial bin. The dashed lines show the $5^{th}$ and $95^{th}$ quantile of the range of 
radial variation of the average values of $\Delta \log M_{*}$ and $\Delta z$. This range is obtained by randomly 
shuffling the galaxies' projected distance within the full range of projected distances of the group sample while keeping the 
values of $\Delta \log M_{*}$ and $\Delta z$ the same. These ranges provide a measure of the statistical significance of the observed 
radial variation. Note that the error bars for each galaxy are always smaller than 0.1 in $\Delta \log M_{*}$ and 0.001 in $\Delta z$ 
since the control galaxies are selected to have smaller stellar mass and redshift difference than these values.

The radial variation of the average stellar mass compared to the control sample is within 0.01 dex 
which corresponds to less than 3\% difference in stellar mass. The radial variation of redshift difference is within 0.0002. 
The stellar masses and redshifts of the group galaxies do not show any systematic radial variation with 
respect to the control sample, which ensures that any measured systematic trends as a function of projected group centric distance  
are not driven by the radial variation of these properties.

\section{THE INFLUENCE OF THE GROUP ENVIRONMENT ON THE \hi\ CONTENT IN GALAXIES}\label{result} 

The goal of this work is to study the impact of environment on the \hi\ gas content of galaxies in groups and clusters. 
To examine the impact of environment, we compare the gas content of galaxies within groups and clusters to that in field galaxies 
(i.e., the control sample) as a function of projected distance from the group center.

\subsection{Radial Distribution of \hi\ Gas-to-Stellar Mass Ratio as a Function of Halo Mass} \label{result:gasfraction}

Figure \ref{fig:fgas_dist1} shows $\Delta \mbox{log} (M_{\mbox{\tiny HI}}/M_{*})$ as a function of projected distance from the 
group center normalized by $R_{180}$. The value $\Delta \log (M_{\mbox{\tiny HI}}/M_{*})$ is defined as:

\begin{equation}
\Delta \log (M_{\mbox{\tiny HI}}/M_{*}) \equiv \log (M_{\mbox{\tiny HI}}/M_{*})_{group} - \log <(M_{\mbox{\tiny HI}}/M_{*})_{control}>.
\end{equation}
where $<(M_{\mbox{\tiny HI}}/M_{*})_{control}>$ is the average value of \hi\ gas-to-stellar mass ratio of the control sample.  

The three panels show the results for three different halo mass bins, $10^{12.5} h^{-1}$ \msun $< M_h < 10^{13.35} h^{-1}$ \msun , 
$10^{13.35} h^{-1}$ \msun $< M_h < 10^{13.85} h^{-1}$ \msun\ and $ 10^{13.85} h^{-1}$ \msun $< M_h < 10^{15.0} h^{-1}$ \msun. 
The halo mass bins are chosen so that there are similar number of galaxies ($\approx 120-150$) in each bin and 
so that the variation in the number of galaxies across the radial bins is small (i.e., the standard deviation of the number of galaxies 
in the 6 radial bins is $\approx 50-58$\% of the average number of galaxies). The results presented here do not change significantly 
as the boundaries between the bins are shifted from $10^{13.30} h^{-1}$ \msun\ to $10^{13.40} h^{-1}$ \msun\ and from 
$10^{13.80} h^{-1}$ \msun\ to $10^{13.90} h^{-1}$ \msun.
The error bars and symbols in this figure are the same as in Figure \ref{fig:mstar_dist}.
The black points connected by a solid line show the average $\Delta \mbox{log} (M_{\mbox{\tiny HI}}/M_{*})$ in each bin of projected 
distance from the group center.

Overall there is no statistically significant variation in the average $\Delta \log (M_{\mbox{\tiny HI}}/M_{*})$ as a function 
of the projected group centric distance. However, there is a hint of a decreasing $\Delta \log (M_{\mbox{\tiny HI}}/M_{*})$ 
for galaxies in the central region ($d/R_{180}<0.2$) of the group halos in the most massive halo bin \footnote{These massive halos were 
identified with some of known galaxy clusters in the Abell catalog \citep{abell_etal_1989}, including A160, A996, A1185, A1656 (in Coma), 
A1795, A1797, A2040, A2052, A2063, A2147 (in Hercules) and A2592.}. The average $\Delta \log (M_{\mbox{\tiny HI}}/M_{*})$ in 
the smallest projected distance bin is slightly smaller than the range enclosed by the $5^{th}$ and $95^{th}$ quantile of the variation 
but the deviation is small so it can only hint at the impact of the most massive halos on the gas content of galaxies near their centers.

\subsection{Radial Distribution of Detection Fraction as a Function of Halo Mass} \label{result:detection}

\begin{figure}
\centering
\epsfig{figure=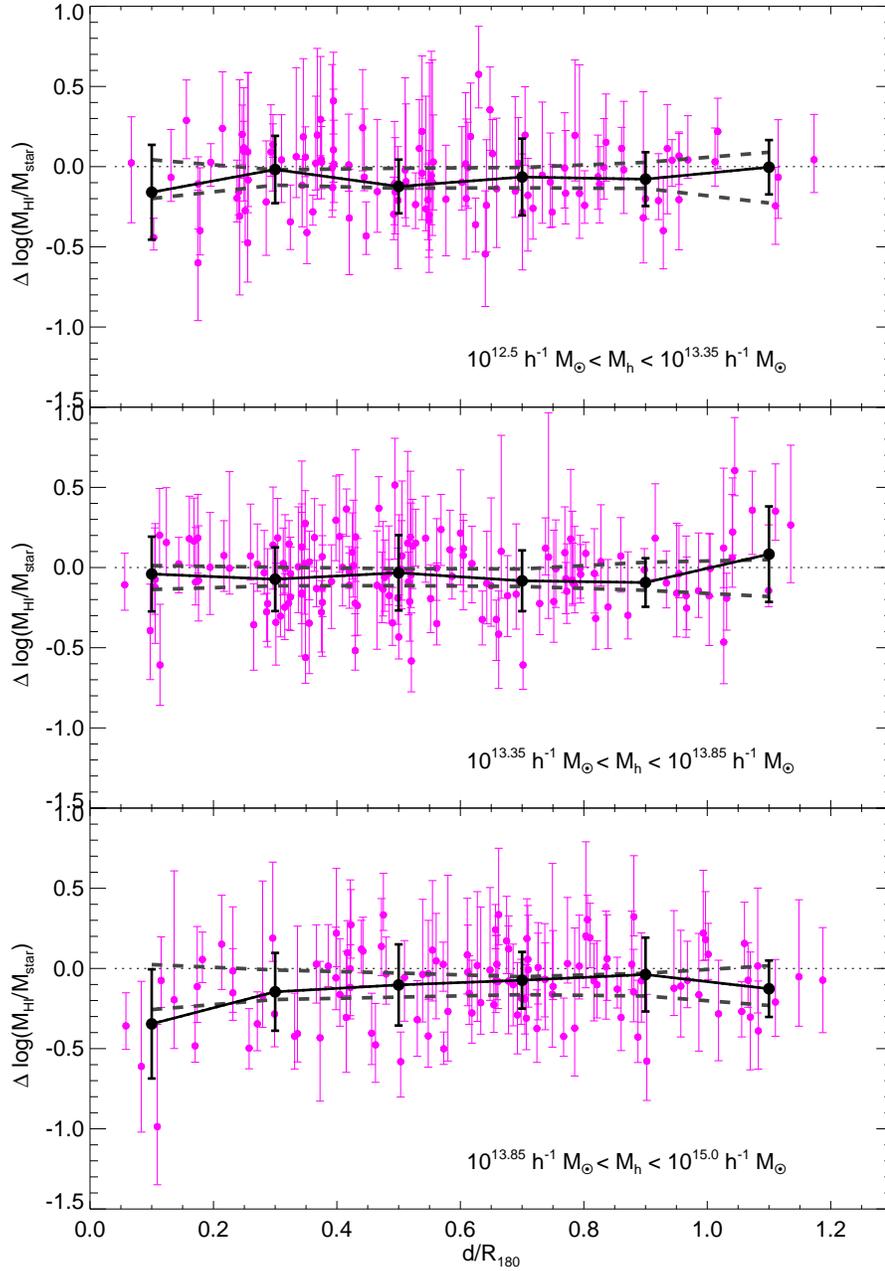, scale=0.70}
\vspace{0.3cm}
\caption{The difference in \hi\ gas-to-stellar mass ratio between galaxies in the group sample and the mean 
of the corresponding 8 control galaxies, as a function of the projected group centric distance for different halo mass bins.
Symbols and lines are the same as those in Figure~\ref{fig:mstar_dist}. 
}
\label{fig:fgas_dist1}
\end{figure}

The lack of a statistically significant result in Figure \ref{fig:fgas_dist1} may be driven by the selection of only the most gas-rich 
galaxies by ALFALFA. Therefore, we investigate the fraction of galaxies detected by ALFALFA as function of projected distance from the 
group center. For this analysis we compute the detection fraction for the group galaxies and compare it with the detection fraction of 
the associated controls. In practice, we define ``detection ratio" as the ratio of the fraction of group galaxies detected at a given 
distance from the group center to the fraction of the associated control galaxies that are detected:

\begin{equation}
\mbox{Detection Ratio} \equiv \frac{(N_{det}/N_{tot})_{group}}{(N_{det}/N_{tot})_{control}}
\label{eq:detect}
\end{equation}
where $N_{det}$ is the number of galaxies detected by the ALFALFA survey and $N_{tot}$ is the total number of SDSS galaxies in the ALFALFA 
survey region.

For each bin of projected group centric distance, we compute the numerator $(N_{det}/N_{tot})_{group}$ 
using all group galaxies that meet the selection criteria defined in Section \ref{sample:group}. Then, for each SDSS group galaxy (both detected
and not detected by ALFALFA) used for computing the numerator $(N_{det}/N_{tot})_{group}$, we identify a set of 8 isolated 
control galaxies (number of group members equals one in the group catalog) with similar stellar mass and redshift ($\Delta M_{*} < 0.1$ dex and $\Delta z < 0.001$).
The denominator in Equation \ref{eq:detect} is the fraction of the ALFALFA detected control galaxies in the projected group centric distance bin.
The denominator provides a normalization of the detection fraction that is sensitive to the mass and distance of the group galaxies 
in each bin. For the control sample, $d/R_{180}$ and halo mass do not have a physical meaning. However, each control galaxy is associated with an 
individual galaxy that does have a projected separation from the group center so each control is associated with a particular projected 
separation through its associated galaxy.

Because there are different ways to select a control sample (e.g., the single best match like in \citet{ellison_etal_2015} or 
allowing control galaxies to be selected only once), we performed this analysis using different selection criteria for the control sample 
and find no significant difference in the results. Therefore, we present the results for 8 controls, consistent with the analysis from 
Section \ref{result:gasfraction}.

The upper left-hand panel of Figure \ref{fig:detect_dist} shows the detection ratio for 
galaxies in group halos as a function of the group-centric distance for the same three halo mass bins as in Figure \ref{fig:fgas_dist1}. 
Error bars are Poisson errors. The average detection fraction for the control galaxies is 27\% and there is little variation among the three halo 
mass sub-samples (as expected since these are all isolated galaxies and are only associated with a halo mass bin because of the galaxies for which they 
serve as a control). When broken down further, there is a 7\% variation in the detection fraction of the control samples across the radial bins with 
which they are associated (the higher variance is due to the smaller number of control galaxies associated with each radial bin).

For all halo masses there is a decrease in the detection ratio closer to the center of the groups. 
The decrease in detection ratio is significantly stronger for galaxies in the most massive halo bin reaching only 6\% of the detection fraction of 
the control sample in the smallest projected distance bin at $d/R_{180}=0.1$. 

We can only measure the \textit{projected} distance of galaxies with respect to the group center and so in each projected radius bin
there are foreground or background ``interlopers'', group galaxies at significantly larger real separations from the group center than their projected separations. 
Therefore, the detection fraction in a given projected separation bin is an upper limit
because these interlopers are further from the group center and thus more similar to the field galaxy population. 
To estimate the population of these interlopers in each bin, we use the projected mass profile from \citet{lokas_and_mamon_2001} which is
based on an NFW profile \citep{nfw1997} with concentration parameter $c=7$. At projected distances of  $d/R_{180}=0.1, 0.3, 0.5, 0.7, 0.9$, 
and $1.1$, these model predicts interloper fractions of 0.33, 0.16, 0.11, 0.07, 0.06, and 0.05 respectively. 

In the inner-most radial bin $d/R_{180}=0.1$, the numbers of group galaxies detected by ALFALFA are 16, 25, 9 and the total number of group galaxies are 
362, 366, 486, for the lowest, middle, and highest halo mass bins respectively, which corresponds to detection fractions of 4\%, 7\% and 2\% 
respectively. If one third of galaxies in the inner-most radial bin are interlopers (120, 120 and 161 for the each halo mass bin), 
then even a modestly higher detection rate for these galaxies that only appear to be in the group because of projection effects would account for many of 
the detections. More relevant than the specific numbers, which are highly affected by small number statistics, is that these low detection rates in the 
centers of groups and clusters are likely to be upper limits because of these interlopers. In addition, the sample has been selected to avoid the densest 
groups (no more than one galaxy in the beam of the radio telescope) where some of these effects are likely to be even more severe.

\subsubsection{Effect of Galaxy Stellar Mass} \label{result:detection_mstar}

\begin{figure}
\centering
\epsfig{figure=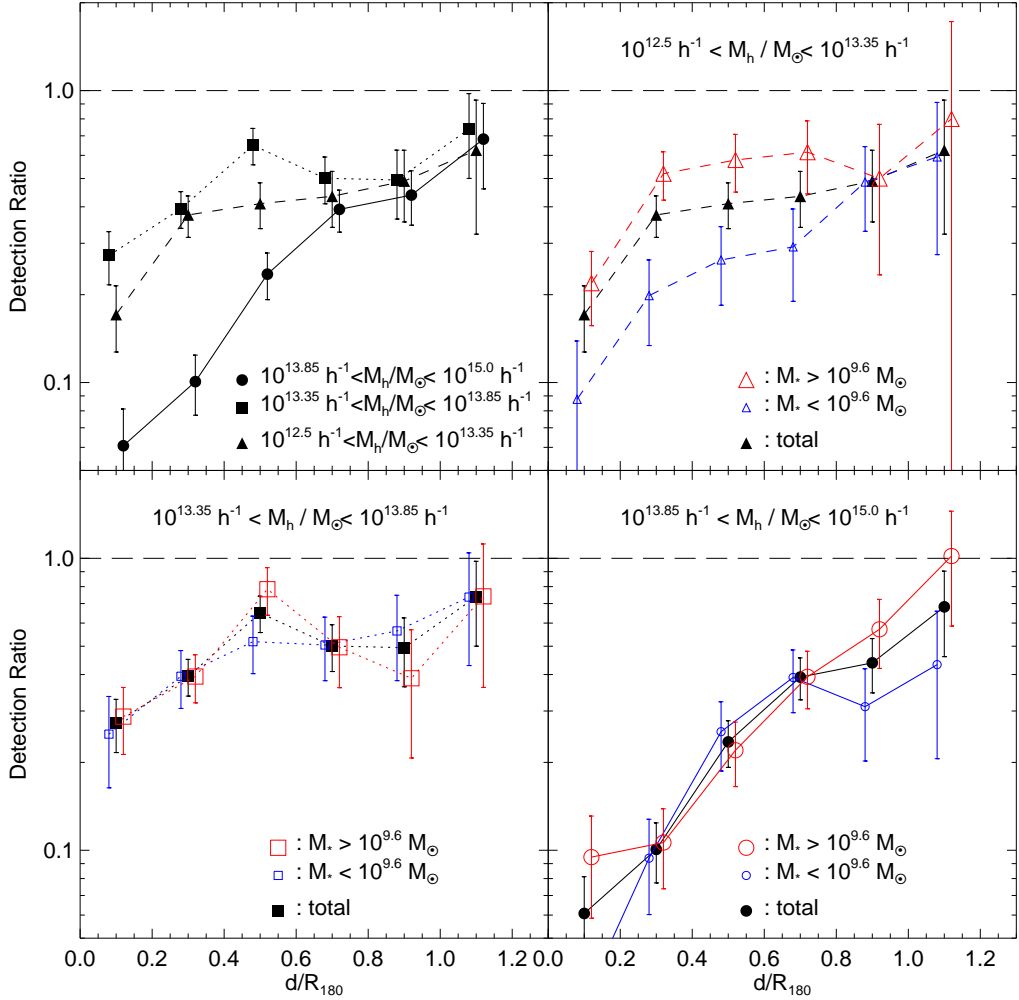, scale=0.70}
\vspace{0.3cm}
\caption{\textit{Upper left panel:} The variation of the detection ratio as a function of the projected group centric distance for different 
halo mass bins. \textit{Other panels:} The detection ratio for each halo mass bin with the detection ratio for two groups of galaxies 
divided by galaxy stellar mass. Open points connected by red and blue lines show these relationships for massive galaxies 
($M_{*}>10^{9.6} M_{\odot}$) and low mass galaxies ($M_{*}<10^{9.6} M_{\odot}$). Error bars are Poisson errors. 
The detection ratio decreases toward group center and the decrement is most dramatic in large mass halos.
}
\label{fig:detect_dist}
\end{figure}

The differences in the depth of the gravitational potential well mean that the impact of environment on the \hi\ in a galaxy may be 
correlated with galaxy mass.
The three additional panels of Figure \ref{fig:detect_dist} show the detection ratio broken down by the galaxy stellar mass
for each halo mass bin. The galaxies are divided into high (larger red symbol) and low (smaller blue symbol) stellar 
mass systems at a mass of $10^{9.6}$\msun\ which roughly corresponds to the median of the galaxy stellar mass distribution in the group sample. 
The detection ratio for low stellar mass galaxies ($\sim 0.25$) is significantly less than the ratio for high stellar mass galaxies ($\sim 0.55$) inside the 
virial radius ($d/R_{180} < 0.8$) for the lowest halo mass bin (upper right-hand panel of Figure \ref{fig:detect_dist}). The impact of the stellar mass of 
the galaxy on the detection ratio becomes less significant with increasing halo mass. 
For the highest mass halos, the detection ratio for low mass galaxies ($M_{*}<10^{9.6}$\msun) just beyond the virial radius 
is only 43\% of that in the field, corresponding to a $2\sigma$ deviation. Although the statistical significance is not large, 
this result may indicate that small galaxies in the outskirts 
of massive halos are already gas poor compared to the field galaxies with similar stellar mass and redshift.

\subsubsection{Effect of Galaxy Type: \sersic\ index} \label{result:detection_sersicn}

\begin{figure}
\centering
\epsfig{figure=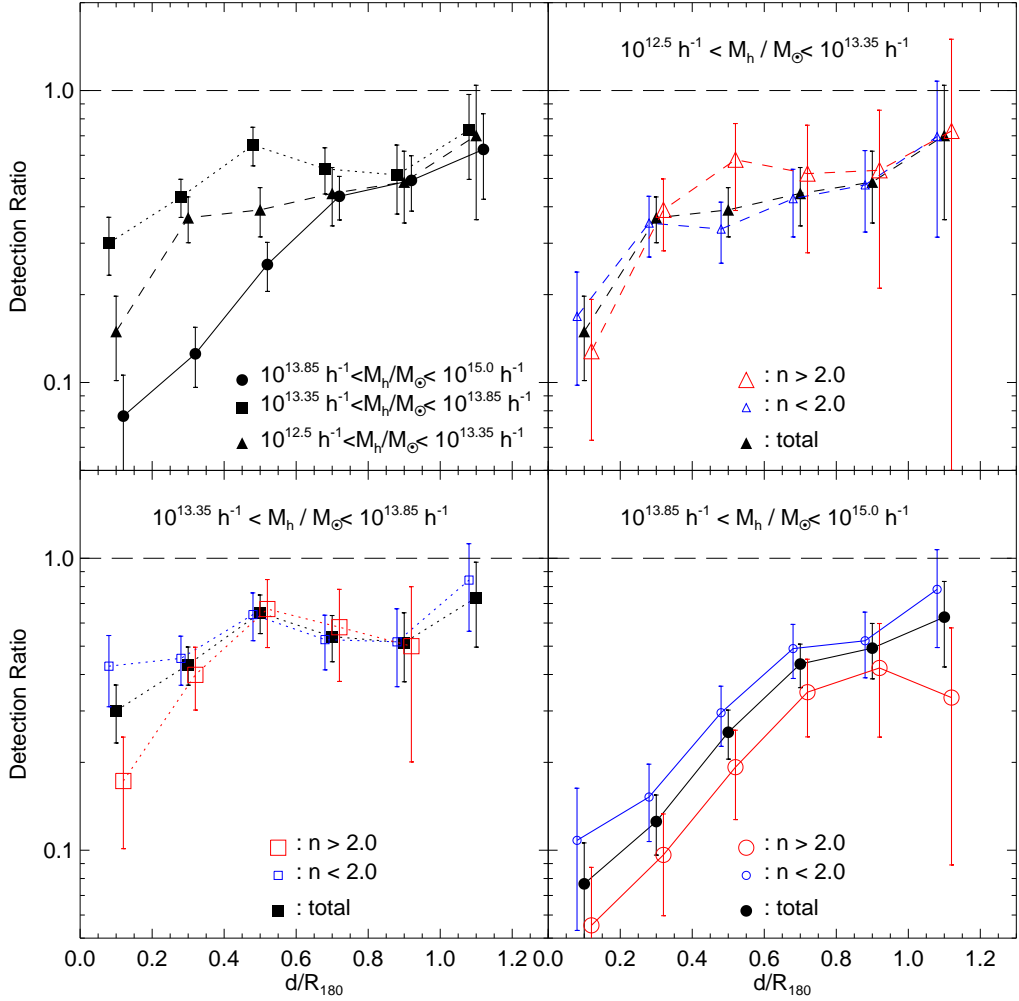, scale=0.70}
\vspace{0.3cm}
\caption{\textit{Upper left panel:} The variation of the detection ratio as a function of the projected group centric distance for different 
halo mass bins. \textit{Other panels:} The detection ratio for each halo mass bin for galaxies 
separated by galaxy \sersic\ index. Open points connected by red and blue lines show these relationships for early type galaxies 
($n>2$) and late type galaxies ($n<2$). Control samples for each early and late type galaxy are selected from the pool of
isolated field galaxies with $n>2$ and $n<2$ respectively.
Error bars are Poisson errors. The detection ratio decreases toward group center for both
early and late type galaxies.
}
\label{fig:detect_dist_sersicn}
\end{figure}

The morphology-density relation \citep{dressler_1980} implies that there are more early-type galaxies towards the centers of 
galaxy clusters \citep[e.g.,][]{blanton_etal_2005}. To examine whether the decreasing detection ratio in Figure \ref{fig:detect_dist} is due to decreasing 
amounts of gas in the galaxies or to different galaxy morphologies as a function of position in the group the three 
additional panels of Figure \ref{fig:detect_dist_sersicn} show the detection ratio broken down by the galaxy \sersic\ index
for each halo mass bin. The galaxies are divided into high (larger red symbol) and low (smaller blue symbol) \sersic\ index 
systems at $n=2.0$ and then control sample for each galaxy with $n>2.0$ (and $n<2.0$) is drawn from the pool of isolated galaxies 
with $n>2.0$ (and $n<2.0$), using the same stellar mass and redshift criteria that we used. 
This cut in \sersic\ index provides a very clean sample of late-type galaxies \citep[][]{maller_etal_2009} 
that can be examined for gas depletion. However, the $n>2$ sample is more mixed and thus more difficult to interpret.
We also controlled \sersic\ index when making control sample by restricting \sersic\ index difference between group galaxy and 
controls (i.e., $\Delta n = 2$), however found that the result has not been changed qualitatively.
The detection ratio for both low and high \sersic\ index galaxies decreases toward the group center indicating that even in the disk galaxies gas is 
being depleted towards the center of the groups.

\subsection{Halo Mass Dependent Ram Pressure Stripping}\label{result:criterion}

The strongly decreasing detection ratio of galaxies towards the center of the most massive halos compared with that of the lower mass 
halos suggests a gas depletion process that is dependent on halo mass. The \hi\ gas depletion seen in galaxies that reside in massive clusters is best 
explained by ram pressure stripping \citep[e.g.,][]{boselli_and_gavazzi_2006,cortese_etal_2011,roediger_2009,vollmer_etal_2001}. 
Some galaxies in groups also represent strong candidates for ram pressure stripping 
\citep[e.g.,][]{bureau_and_carignan_2002,mcconnachie_etal_2007,rasmussen_etal_2006,sengupta_etal_2007} although it is harder to distinguish the 
impact of ram pressure stripping from that of tidal interactions in these lower mass halos \citep[e.g.,][]{kilborn_etal_2009,rasmussen_etal_2008}.
The decrease in the detection fraction for galaxies with small ($n<2$) \sersic\ indices implies that gas is being removed from 
the system regardless of the galaxy type and that mergers that can transform galaxies from late to early types are not the dominant reason 
for this result. This gas removal can take two forms in these environments, ram pressure stripping (removal of cold gas from the disk of the galaxy) and starvation 
(removal of hot gas from the galaxy halo). Since starvation can not be probed directly with these \hi\ observations, we focus our discussion on 
ram pressure stripping. The simple criterion for ram pressure stripping first proposed by \cite{gunn_and_gott_1972} is:

\begin{equation}
\label{eq:gg}
\rho_{\mbox{\tiny ICM}} v^2 \ge 2\pi G \Sigma_{*} \Sigma_{\mbox{\tiny ISM}}
\end{equation}
where $\rho_{\mbox{\tiny ICM}}$ is the density of the hot intracluster medium (ICM), $v$ is 
the speed of infalling galaxy, $\Sigma_{*}$ is the galaxy stellar surface 
density, and $\Sigma_{\mbox{\tiny ISM}}$ is the galaxy interstellar medium (ISM) surface density. 
This criterion implies that gas stripping occurs if the ram pressure (i.e., force per unit area) is greater than 
the gravitational binding force per unit area in the ISM. 

We provide a simple analytic calculation based on the Gunn and Gott (GG) criterion to examine the effect of halo mass on ram 
pressure stripping. The GG criteria has previously been modified to be expressed in terms of specific observables: for example, 
IGM electron density, a component of the galaxy velocity relative to the group, central stellar velocity dispersion of 
the galaxy, and \hi\ gas density of the galaxy \citep[e.g.,][]{freeland_and_wilcots_2011,grcevich_and_putman_2009,grebel_etal_2003}. 
To study the impact of halo mass on ram pressure stripping for a range of galaxy stellar masses we use the virial theorem and galaxy 
scaling relations to rewrite the left-hand side of the GG criterion to have explicit dependence on group halo mass and ICM density 
and the right-hand side to depend only on galaxy stellar mass (see Appendix A for details). With these changes the GG criterion 
depends on the mean electron number density in groups or clusters, the halo mass and the galaxy stellar mass:

\begin{equation}
\label{myeq:rampressure2}
\left(\frac{n}{10^{-3} \mbox{cm}^{-3}}\right)\left(\frac{M_h}{10^{14} h^{-1} M_{\odot}}\right)^{2/3} \gtrsim
1.24 \left(\frac{M_{*}}{10^{10} M_{\odot}}\right)^{0.58} \left(1+0.25\times\frac{M_{*}}{10^{10} M_{\odot}}\right)^{-1}.
\end{equation}

Massive halos have high ICM densities \citep[e.g.,][]{cavagnolo_etal_2009} due 
to their strong gravitational potentials while small mass halos tend to have 
low ICM densities. The halo mass is also related with the X-ray luminosity through 
an observational scaling relation \citep[e.g.,][]{arnaud_etal_2007}.
Therefore, the virial theorem and X-ray scaling relation imply that the mean 
electron number density of the ICM, the halo mass, and the kinetic temperature 
of the ICM are related by (see Appendix A for details):

\begin{equation}
\label{myeq:xrayscale}
\left(\frac{n}{10^{-3} \mbox{cm}^{-3}}\right) \approx
0.68 h^3 \left(\frac{T}{\mbox{keV}}\right)^{-1} \left(\frac{M_{h}}{10^{14} h^{-1} M_{\odot}}\right)^{0.824}.
\end{equation}

By combining these two relations and neglecting the redshift dependence, one can determine where the GG criterion is valid 
in the group halo mass and galaxy stellar mass plane for a given ICM temperature. Figure \ref{fig:mhalo_rps} 
shows the distribution of the group sample (black points) in the halo mass and galaxy stellar mass plane for galaxies within the virial 
radius of their group halo (where ram pressure is expected to play a role) with black lines 
representing Equation \ref{myeq:rampressure2} for ICM temperatures ranging from 1 keV to 8 keV. 
The colored regions show the average $\Delta \mbox{log} (M_{\mbox{\tiny HI}}/M_{*})$ of galaxies in each grid region weighted 
by their errors in $\Delta \mbox{log} (M_{\mbox{\tiny HI}}/M_{*})$. Regions with 0-1 galaxy have not been colored.
The region above the black line for a given ICM temperature is where ram pressure wins over the surface gravity of the galaxy. 
Galaxies in halos with $M_h>10^{13.85} h^{-1}$\msun, are mostly located in the region where ram pressure wins, for a range of 
typical ICM temperatures (1-8 keV). Alternatively, galaxies in lower mass halos generally reside in the region where galaxy surface 
gravity wins over ram pressure. The galaxies with the lowest $\Delta \mbox{log} (M_{\mbox{\tiny HI}}/M_{*})$ reside in the two bins 
where ram pressure stripping is expected to play the largest role.

Figure \ref{fig:mhalo_rps} implies that for the stellar mass range of the galaxies in our 
sample $10^{8.4} M_{\odot} \le M_{*} \le 10^{10.6} M_{\odot}$, and typical ICM temperature (1-8 keV), ram pressure is efficient 
for halo masses larger than $\approx 10^{13-14} h^{-1}$\msun, which explains the decreasing detection 
ratio toward the group center prominently seen in the most massive halo bin $M_h>10^{13.85} h^{-1}$\msun.
Similarly \citet{catinella_etal_2013} find that massive galaxies ($M_{*}>10^{10}$\msun) in halos with $M_h>10^{13-14}$ \msun\ show a 
decrement of \gapp0.4 dex in \hi\ gas-to-stellar mass ratio relative to the similar galaxies in lower density environments. Using 
semi-analytic galaxy formation models, \citet{fabello_etal_2012} also infer that \hi-deficient galaxies are more common in dark matter 
halos with $M_h>10^{13}$ \msun. The modified ram pressure stripping criterion that depends on halo mass is used here to show that 
ram pressure stripping can be effective in depleting the \hi\ for wider
range of stellar mass $10^{8.4} M_{\odot} \le M_{*} \le 10^{10.6} M_{\odot}$.

\section{DISCUSSION}\label{discussion}

\subsection{Gas Removal in Group and Cluster Halos}

Using the SDSS and ALFALFA surveys, we have examined the evidence for ram pressure stripping of \hi\ gas in groups and clusters. 
We find no statistically significant evidence for a change in the galaxy \hi\ gas-to-stellar mass ratio as a function of projected distance from 
the group center in this sample despite the existing evidence for gas depletion in 
clusters \citep{haynes_etal_1984,haynes_etal_1986,solanes_etal_2001}. This result is, however, consistent with the previous \hi\ studies 
of gas depletion in Virgo Cluster galaxies \citep{taylor_etal_2012} and Coma Supercluster galaxies (Abell 1367, \citealp{cortese_etal_2008}). 
The distribution of \hi\ gas-to-stellar mass ratio in galaxies shows a large scatter (e.g., 0.43 dex in \citealp{catinella_etal_2012}) even 
at fixed stellar mass which drives a large standard deviation (approximately 0.4 dex or a factor of 2.5 as seen in Figure \ref{fig:sigmatest}) 
in $\Delta \mbox{log} (M_{\mbox{\tiny HI}}/M_{*})$. Although, in principle, it is possible to detect a difference
in the average \hi\ gas-to-stellar mass ratio between groups and field with enough statistics, our sample of groups with 8 controls (390) is
not enough to detect the difference. A robust detection of gas depletion in groups will require 
deeper \hi\ surveys that increase the dynamic range of detected gas fractions for a large number of 
galaxies in groups and clusters.

While the direct measurement of gas removal in groups and clusters using ALFALFA detected galaxies alone is not possible 
without relying on deeper \hi\ survey for a sample with stellar mass threshold (GASS; \citealp{catinella_etal_2010,catinella_etal_2012}) 
or stacking of ALFALFA spectra \citep{fabello_etal_2012}, we confirm the 
results of \cite{hess_and_wilcots_2013} using a sample with matched controls and measure this removal indirectly through the decreasing 
detection ratio, which is a ratio of detection 
fraction for group galaxies to that for isolated galaxies. By controlling our sample for stellar mass and redshift we show 
that this is a real effect not a function of selection effects and sampling. We find that most significant decrease in detection ratio occurs in the 
most massive halos, as expected from ram pressure stripping discussed in Section \ref{result:criterion}. This result is consistent 
with previous studies by \citet{catinella_etal_2013} and \citet{fabello_etal_2012}. Nevertheless, the detection ratio does decrease 
significantly in the inner-most regions of even the lowest mass groups which is consistent with the detection of X-rays in some halos of 
these masses \citep[e.g.,][]{helsdon_and_ponman_2000,mulchaey2000,zabludoff_and_mulchaey_1998}. 

Assuming the IGM temperature to be 1 keV for the lowest mass groups ($M_h<10^{13.35} h^{-1}$\msun) in this study, 
our calculation indicates that ram pressure should only be effective for galaxies with stellar masses below $10^8$ \msun\ yet all of the galaxies 
in this sample are more massive than this limit and still show a decreasing detection ratio towards the group center. Either the IGM 
temperature in these low mass groups is lower than assumed \citep[$<1$ keV, e.g.,][]{helsdon_and_ponman_2000,trinchieri_et_al_2012} implying 
a larger mean IGM density than $8 \times 10^{-5} \mbox{cm}^{-3}$ from Equation \ref{myeq:xrayscale} which is similar to the value for poor 
groups \citep[$8.9 \times 10^{-5} \mbox{cm}^{-3}$, ][]{zabludoff_and_mulchaey_1998} enough for ram pressure stripping of dwarf 
galaxies \citep[e.g.,][]{bureau_and_carignan_2002} or a process other than ram pressure stripping is 
responsible for removing the gas in these low mass halos \citep[e.g.,][]{rasmussen_etal_2012}. 

Using a large sample ($N \approx 23000$) of galaxies with \hi\ mass inferred from optical photometry, \cite{zhang_etal_2013} 
investigate the effect of galaxy stellar mass on the \hi\ content of galaxies in groups and clusters and find that 
lower stellar mass galaxies show a stronger radial decrease in \hi\ gas content toward the brightest cluster galaxy and that
this trend becomes more significant for galaxies with lower stellar surface density. While we can not probe the effect of stellar 
surface density for given galaxy stellar mass due to small sample size, Figure \ref{fig:detect_dist} shows that for the lowest mass halos it is 
only the low mass galaxies that are significantly stripped while in the high mass halos all of the galaxies are likely to be impacted. 
Our result that the halo mass impacts the degree of gas depletion contrasts with those of \cite{zhang_etal_2013} who find no difference 
as a function of group velocity dispersion as a proxy for halo mass. This difference may be related to the large dispersion in the 
velocity dispersion - halo mass relation \citep[e.g.,][]{weinmann_etal_2010} and/or the additional impact of tidal 
stripping on the gas in their sample galaxies \citep{zhang_etal_2013}. For this study galaxies with neighbors 
within 1.75 \arcm\ (which means no pair galaxies within 20 kpc at $z=0.01$ and 105 kpc at $z=0.055$) have been excluded. 
Many of the excluded galaxies with small velocity separations are likely to be undergoing tidal interactions as suggested 
by \cite{scudder_etal_2012}. Therefore, this sample should be significantly less strongly impacted by tidal interactions 
than \cite{zhang_etal_2013} and should more clearly show the impact of ram pressure stripping on these galaxies.

\subsection{Effect of Pre-processing}

In hierarchical models, the gas content in galaxies decreases during interactions within small groups before they enter a large 
halo \citep[e.g.,][]{fujita_2004}. Observing this pre-processing of small groups is challenging because the current merger rate 
is considerably lower than in the past \citep{gottlober_etal_2001}. However, there is supporting evidence for 
pre-processing revealed, for example, by the star formation quenching as a function of galaxy environment in and around the Coma 
Supercluster \citep[][]{cybulski_etal_2014}; the distribution of post-starburst galaxies in sub-structures of galaxy 
clusters \citep[][]{mahajan_2013}; and the ionized gas distribution in compact groups of galaxies \citep{cortese_etal_2006}. 

The detection ratio of galaxies at the largest group-centric distances (i.e., beyond the virial radius) may provide a measurement of 
how environment has affected the systems prior to their falling into the group. The upper left panel of Figure \ref{fig:detect_dist} shows 
with marginal (1.5$\sigma$) statistical significance that the \hi\ detection fraction in massive halos ($M_h > 10^{13.85} h^{-1}$\msun) 
is $68$\% of the detection fraction for the control sample even beyond the virial radius (i.e., $d/R_{180} \approx 1$). 
The effect is even more strongly pronounced for the lowest mass galaxies as shown in the lower right panel of Figure \ref{fig:detect_dist}, 
where the detection fraction of the lowest mass galaxies is less than that of the control sample by 43\% corresponding to a $2\sigma$ difference. 

While not definitive because of the low statistical significance, this depression in the detection fraction for small galaxies in massive halos 
provides a tantalizing indication that these small galaxies are already \hi\ deficient when they fall into the massive halos, as also suggested 
by \citet{jaffe_etal_2012} based on the \hi\ measurement of compact group. This trend of gas deficiency at the outskirt of massive halos supporting 
pre-processing scenario is predicted by semi-analytic model of mass accretion \citep[e.g.,][]{mcgee_etal_2009} and hydrodynamic simulation of \hi\ gas 
content \citep[e.g.,][]{rafieferantsoa_etal_2014}, and also implied by recent observational studies using statistical 
sample \citep[e.g.,][]{hess_and_wilcots_2013,hou_etal_2014}.

In considering the significance of this depression at large group centric distance and the other results presented here, it is important to consider 
whether the sample selection criteria (excluding groups with fewer than 5 members, excluding galaxies with close neighbors, and excluding galaxies with 
fewer than 8 controls) might bias these results. While bias can not be ruled out, these selection criteria should not influence these results within 
the bounds of the sample (i.e., we can not evaluate the impact of the group environment on the smallest and most compact groups) with the exceptions 
of the close neighbors. The removal of galaxies with close neighbors will preferentially eliminate interacting pairs. Interacting pairs, however, have 
not shown any significant decrease or increase in their detection fraction (Fertig et al. 2015 submitted) implying that the impact of removing galaxies 
with close neighbors will not introduce a significant bias to the detection ratio estimate.

\section{CONCLUSIONS}\label{conclusion}

We have investigated the \hi\ gas content in groups and clusters over a wide range of halo masses
and the results of this study are: 

\begin{itemize}

\item There is no statistically significant variation in \hi\ gas-to-stellar mass ratio with projected distance from the group center in the sample studied here. 

\item The detection ratio of galaxies decreases towards the center of groups spanning the range of masses studied here with the most significant 
decrease occurring towards the centers of the most massive halos. 

\item We frame the GG criteria in terms of halo mass and stellar mass and find that the change in the radial distribution of 
detection ratio with respect to halo mass is qualitatively consistent with ram pressure stripping while the decrease in the detection ratio for galaxies 
with $M_{*}<10^{9.6}$ \msun\ in the lowest mass halos is not consistent with ram pressure stripping unless the temperature of the IGM 
is lower than assumed (i.e., $< 1$ keV). 

\item The detection ratio for low ($M_{*}<10^{9.6}$\msun) stellar mass galaxies just beyond the virial radius of the most massive groups 
is 2$\sigma$ below unity, which could be due to pre-processing of these galaxies before they fall into the larger groups. 

\end{itemize}

The lack of a significant detection of \hi\ depletion in groups and clusters, indicates that ALFALFA is not sufficiently deep to measure 
this effect because of the large galaxy to galaxy dispersion in gas content. Even deeper \hi\ surveys like the Arecibo Galaxy Environment 
Survey (AGES) do not find strong evidence for \hi\ deficiency increasing toward the center of the Virgo \citep{taylor_etal_2012} and 
Coma/A1367 \citep{cortese_etal_2008} clusters. Explicitly measuring the amount of environment dependent gas depletion will require significantly 
deeper large statistical samples which might have to wait for the full SKA \citep{catinella_etal_2013}.

\acknowledgments

This research was supported by NSF grant AST-000167932. We appreciate the valuable comments from the anonymous referee. 
We thank Xiaohu Yang for providing his SDSS DR7 halo group catalog. We thank Derek Fertig and Jessica O'Connor for useful discussions 
of the ALFALFA data and acknowledge useful comments from Jacqueline van Gorkom. We also acknowledge the work of the entire ALFALFA 
collaboration team in observing, flagging, and extracting the catalog of galaxies used in this work.

Funding for the SDSS and SDSS-II has been provided by the Alfred P. Sloan Foundation, the Participating Institutions, 
the National Science Foundation, the U.S. Department of Energy, the National Aeronautics and Space Administration, 
the Japanese Monbukagakusho, the Max Planck Society, and the Higher Education Funding Council for England. 
The SDSS Web Site is http://www.sdss.org/.

The SDSS is managed by the Astrophysical Research Consortium for the Participating Institutions. The Participating Institutions 
are the American Museum of Natural History, Astrophysical Institute Potsdam, University of Basel, University of Cambridge, 
Case Western Reserve University, University of Chicago, Drexel University, Fermilab, the Institute for Advanced Study, 
the Japan Participation Group, Johns Hopkins University, the Joint Institute for Nuclear Astrophysics, the Kavli Institute for 
Particle Astrophysics and Cosmology, the Korean Scientist Group, the Chinese Academy of Sciences (LAMOST), Los Alamos National 
Laboratory, the Max-Planck-Institute for Astronomy (MPIA), the Max-Planck-Institute for Astrophysics (MPA), New Mexico State 
University, Ohio State University, University of Pittsburgh, University of Portsmouth, Princeton University, the United States 
Naval Observatory, and the University of Washington.

\appendix
\section{A simple form of the Gunn and Gott criterion depending on halo mass} \label{appdx}
A simple criterion for ram pressure stripping first proposed by \cite{gunn_and_gott_1972} is
\begin{equation}
\rho_{\mbox{\tiny ICM}} v^2 \ge 2\pi G \Sigma_{*} \Sigma_{\mbox{\tiny ISM}}
\end{equation}
where $\rho_{\mbox{\tiny ICM}}$ is the density of hot intracluster medium (ICM), $v$ is the speed of infalling galaxy,
and $\Sigma_{*}$ and $\Sigma_{\mbox{\tiny ISM}}$ is the density of galaxy stellar mass and interstellar medium (ISM) respectively. 
Using a simple assumption that the galaxy stellar mass ($M_{*}$) and ISM mass ($M_{\mbox{\tiny ISM}}$) are distributed
along the same characteristic radius $r_g$, we can rewrite this criterion as
\begin{equation}
\label{eq:app1}
\left(\frac{n}{10^{-3} \mbox{cm}^{-3}}\right)\left(\frac{\sigma}{1000 \mbox{km s}^{-1}}\right)^2 \gtrsim
0.37 \left(\frac{M_{*}}{10^{10} M_{\odot}}\right)^2 \left(\frac{M_{\mbox{\tiny ISM}}}{M_{*}}\right) \left(\frac{r_g}{10 \mbox{kpc}}\right)^{-4}
\end{equation} where $n$ is the mean electron number density in a galaxy cluster such that: $\rho_{\mbox{\tiny ICM}} = n m_{p}$ 
and $\sigma$ is the line of sight velocity dispersion of galaxies in groups and clusters (i.e., $v^2 \approx 3\sigma^2$).

If one defines the halo mass $M_h=M_{180}$ and the halo virial radius $R_{vir}=R_{180}$ using the quantities at an overdensity of 
180 and combines the virial theorem 
$\sigma^2 = \frac{2GM_h}{R_{vir}}$ and Equation \ref{eq:rvir}, one can modify Equation \ref{eq:app1} and the ram pressure 
criterion including halo mass is written as 
\begin{equation}
\left(\frac{n}{10^{-3} \mbox{cm}^{-3}}\right) \left(\frac{M_h}{10^{14} h^{-1} M_{\odot}}\right)^{2/3} \gtrsim
4.23 \left(\frac{M_{*}}{10^{10} M_{\odot}}\right)^2 \left(\frac{M_{\mbox{\tiny HI}}}{M_{*}}\right) \left(\frac{r_g}{10 \mbox{kpc}}\right)^{-4}.
\label{eq:rampressure}
\end{equation} where we replace $M_{\mbox{\tiny ISM}}$ by cold atomic gas mass $M_{\mbox{\tiny HI}}$.

Furthermore, if we adopt the scaling relations between $M_{*}$ and $r_g$ and between $M_{*}$ and $M_{\mbox{\tiny HI}}$,
the physics on the right hand side of the above criterion can be described by galaxy stellar mass $M_{*}$ only.
Among galaxy scaling relations, we use the relation 
$M_{\mbox{\tiny HI}}/M_{*} \approx 0.72 \left(\frac{M_{*}}{10^{10}M_{\odot}}\right)^{-0.86}$ from \cite{masters_etal_2012} 
which finds the similar relation to other studies \citep{catinella_etal_2010,fabello_etal_2011,toribio_etal_2011}
and the relation between $M_{*}$ and $r_g$ derived from the relation between galaxy stellar mass and half-light radius 
for late type galaxies with \sersic\ index $n<2.5$, $\frac{r_{50}}{\mbox{kpc}} \approx 
0.1 \left(\frac{M_{*}}{M_{\odot}}\right)^{0.14} \left(1+\frac{M_{*}}{3.98\times10^{10}M_{\odot}}\right)^{0.25}$
in \cite{shen_etal_2003}. We define $r_g$ as $5 \times r_{50}$ which encloses nearly 100\% and 90\% of the total luminosity from
the \sersic\ profile for $n=1$ and $n=4$ respectively \citep[e.g.,][]{yoon_etal_2011}.
Then we can simplify the Equation \ref{eq:rampressure} as follows.
\begin{equation}
\label{eq:rampressure2}
\left(\frac{n}{10^{-3} \mbox{cm}^{-3}}\right)\left(\frac{M_h}{10^{14} h^{-1} M_{\odot}}\right)^{2/3} \gtrsim
1.24 \left(\frac{M_{*}}{10^{10} M_{\odot}}\right)^{0.58} \left(1+0.25\times\frac{M_{*}}{10^{10} M_{\odot}}\right)^{-1}.
\end{equation}
  
\begin{figure}
\centering
\epsfig{figure=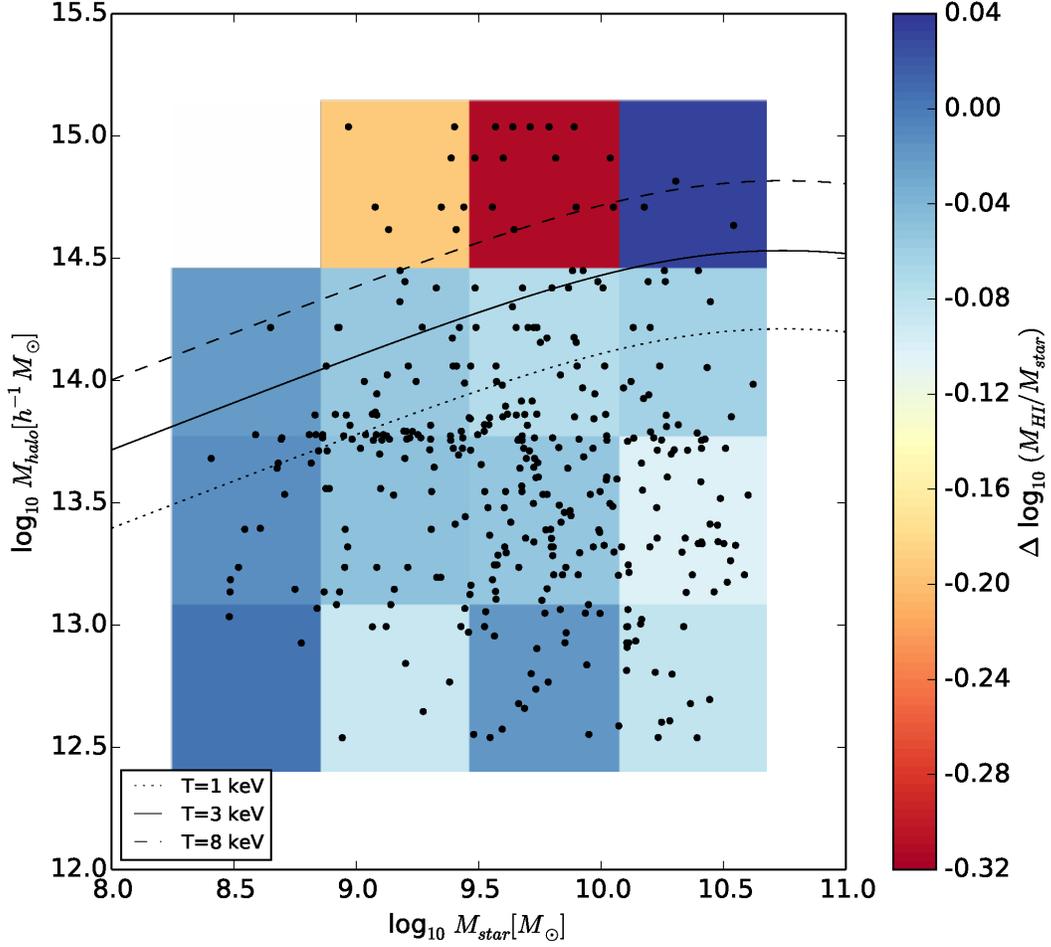, scale=0.80}
\caption{The halo mass-galaxy stellar mass plane that determines the conditions for efficient ram pressure stripping. 
Black lines represent different ICM temperatures based on Equation \ref{myeq:rampressure2} and divide the halo mass-galaxy 
stellar mass parameter space into the regions where ram pressure dominates and where galaxy gravity dominates. Ram pressure stripping 
is efficient in the region above the black lines for a given ICM temperature. The distribution of the group galaxy sample used 
in this study is shown by black points, the colored squares represent the weighted average 
of $\Delta \mbox{log} (M_{\mbox{\tiny HI}}/M_{*})$ in each bin.} 
\label{fig:mhalo_rps}
\end{figure}

On the other hand, soft X-ray emission from galaxy cluster can be used to infer the mean electron number density $n$
of the cluster. One of the robust scaling relations is the $Y_X$-$M_{500}$ relation \citep{arnaud_etal_2007}:
\begin{equation}
h^{2/5}(z) M_{500} \approx 10^{14.40} \left(\frac{M_{g,500} T_X}{2\times10^{14}M_{\odot} keV}\right)^{0.548} h^{-1} M_{\odot}
\end{equation} where $h^2(z) = \Omega_M (1+z)^3 + \Omega_{\Lambda}$, $T_X$ is the kinetic temperature of ICM,
$M_{500}$ is the total mass of a cluster enclosed by the radius $R_{500}$ and $M_{g,500}$ is the mass of hot gas enclosed 
by $R_{500}$. By substituting $M_{g,500}$ with $\frac{4\pi}{3} R^3_{500} n m_p$, 
we express the $n$ given halo mass and temperature at $z=0$ (i.e., $h(z)=1$) as follows
\begin{equation}
\label{eq:xrayscale}
\left(\frac{n}{10^{-3} \mbox{cm}^{-3}}\right) \approx 
0.68 h^3 \left(\frac{T}{\mbox{keV}}\right)^{-1} \left(\frac{M_{180}}{10^{14} h^{-1} M_{\odot}}\right)^{0.824}
\end{equation} where $M_{180}$ is converted from $M_{500}$ \citep[][]{reiprich_etal_2013} by assuming NFW
profile \citep{nfw1997}. By combining Equation \ref{eq:rampressure2} and \ref{eq:xrayscale} and 
using $M_h \approx M_{180}$, one can find a region where ram pressure stripping
becomes efficient for given galaxy stellar mass, halo mass and kinetic temperature of ICM in the halo,
as shown in Figure~\ref{fig:mhalo_rps}.

\bibliography{reference}
\bibliographystyle{mn2e}

\clearpage

\clearpage

\clearpage

\clearpage

\clearpage

\clearpage

\clearpage

\clearpage

\clearpage

\clearpage

\clearpage

\end{document}